\documentclass[12pt]{article}
\usepackage{amssymb,lscape}
\usepackage{amsmath,multirow}
\usepackage{pdflscape}
\usepackage{diagbox}
\usepackage{subfig}
\usepackage[ruled,vlined]{algorithm2e}
\usepackage[normalem]{ulem}

\usepackage{hyperref}

 \usepackage{graphicx}
 \usepackage[curve,matrix,arrow,color]{xy}

\usepackage[usenames,dvipsnames]{color}

\numberwithin{equation}{section}
\def \bes{\begin{eqnarray}}
\def \ees{\end{eqnarray}}
\def \bns{\begin{eqnarray*}}
\def \ens{\end{eqnarray*}}

\newcommand{\ffrac}[2]{\mbox{\footnotesize$\displaystyle\frac{#1}{#2}$}}

\newcommand{\be}{\begin{eqnarray}}
\newcommand{\ee}{\end{eqnarray}}
\newcommand{\non}{\nonumber}
\newcommand{\id}{\mathbb{I}}
\newcommand{\tr}{\mathop{\rm tr}\nolimits}
\newcommand{\diag}{\mathop{\rm diag}\nolimits}
\newcommand{\csch}{\mathop{\rm csch}\nolimits}
\newcommand{\sech}{\mathop{\rm sech}\nolimits}
\newcommand{\qg}{\mathop{U_{q} sl(2)}\nolimits}

\newcommand{\Irr}[1]{\langle#1\rangle}

\begin{document}

\begin{titlepage}
\strut\hfill UMTG--282
\vspace{.5in}
\begin{center}

\LARGE 
Counting solutions of the Bethe equations of the\\
quantum group invariant open XXZ chain\\
at roots of unity\\
\vspace{1in}
\large Azat M. Gainutdinov \footnote{
Fachbereich Mathematik, Universit\"at Hamburg, Bundesstra\ss e 55,
20146 Hamburg, Germany;
DESY, Theory Group, Notkestra\ss e 85, Bldg. 2a, 22603 Hamburg,
Germany, azat.gainutdinov@uni-hamburg.de},
\large Wenrui Hao \footnote{
Mathematical Biosciences Institute, The Ohio State University, 1735
Neil Avenue, Columbus, OH 43210 USA, hao.50@osu.edu},
Rafael I. Nepomechie \footnote{
Physics Department,
P.O. Box 248046, University of Miami, Coral Gables, FL 33124 USA,
nepomechie@physics.miami.edu}\\
and Andrew J. Sommese \footnote{
Department of Applied and Computational Mathematics and Statistics,
University of Notre Dame, 153 Hurley Hall, IN 46556 USA,
sommese@nd.edu}
\\[0.8in]
\end{center}

\vspace{.5in}

\begin{abstract}
We consider the $\qg$-invariant open spin-1/2 XXZ quantum spin chain
of finite length $N$.  For the case that $q$ is a root of unity, we
propose a formula for the number of admissible solutions of the Bethe
ansatz equations in terms of dimensions of irreducible representations
of the Temperley-Lieb algebra; and a formula for the degeneracies of
the transfer matrix eigenvalues in terms of dimensions of tilting
$\qg$-modules.  These formulas include corrections that appear if two
or more tilting modules are 
spectrum-degenerate.
For the XX case ($q=e^{i \pi/2}$), we give explicit formulas for the
number of admissible solutions and degeneracies.
We also consider the cases of generic $q$ and the isotropic ($q
\rightarrow 1$) limit. 
Numerical solutions of the Bethe equations up to $N=8$ are presented.
Our results are consistent with the Bethe ansatz solution being complete.
\end{abstract}

\end{titlepage}

\setcounter{footnote}{0}

\section{Introduction}\label{sec:intro}

The Hamiltonian of the $\qg$-invariant open spin-1/2 XXZ
quantum spin chain with length $N$ is given by \cite{Pasquier:1989kd}
\be
H = \sum_{k=1}^{N-1} \left[ \sigma^x_k \sigma^x_{k+1} +
\sigma^y_k \sigma^y_{k+1} + {1\over 2}( q + q^{-1}) \sigma^z_k
\sigma^z_{k+1}
\right]  - {1\over 2}( q - q^{-1})
\Bigl( \sigma^z_1 - \sigma^z_N \Bigr) \,, \label{Hamiltonian}
\ee
where $\vec\sigma$ are the usual Pauli spin matrices, and $q=e^{\eta}$ is an arbitrary
complex parameter.
This model has been the subject of many investigations (see, for
example \cite{Kulish:1991np, Hou:1991tc, Mezincescu:1991rb,
Mezincescu:1991ag, Alcaraz:1992zc, Destri:1992fa, Juttner:1994sk,
Fridkin:1999qb, Belavin:2000zi, Korff:2007qg,
Gainutdinov:2012mr, Gainutdinov:2012qy}).

This model is solvable by Bethe ansatz \cite{Pasquier:1989kd, Alcaraz:1987uk, Sklyanin:1988yz}: 
the energy eigenvalues are given by
\be
E = 2 \sinh^{2}\eta \sum_{k=1}^{M}
\frac{1}{\sinh(\lambda_{k}-\frac{\eta}{2})\,
\sinh(\lambda_{k}+\frac{\eta}{2})} + (N-1)\cosh\eta \,,
\label{energy}
\ee
where $\{\lambda_{k}\}$ are solutions of the Bethe
equations 
\be
\lefteqn{\sinh^{2N}\left(\lambda_{k} + \frac{\eta}{2}\right)
\prod_{\scriptstyle{j \ne k}\atop \scriptstyle{j=1}}^M
\sinh(\lambda_{k} - \lambda_{j} - \eta)\sinh(\lambda_{k} + \lambda_{j} - \eta)
} \non \\
&&=\sinh^{2N}\left(\lambda_{k} - \frac{\eta}{2}\right)
\prod_{\scriptstyle{j \ne k}\atop \scriptstyle{j=1}}^M
\sinh(\lambda_{k} - \lambda_{j} + \eta)\sinh(\lambda_{k} + \lambda_{j} + \eta) \,, \non \\
&&\qquad k = 1 \,, 2\,, \ldots \,, M \,, \qquad
M = 0\,, 1\,, \ldots\,,  
\Bigl\lfloor\frac{N}{2}\Bigr\rfloor \,,
\label{BAE}
\ee
where 
$\lfloor k \rfloor$ denotes the largest integer not greater than $k$.
This exact solution owes its existence to the fact that
the model is quantum integrable: there are 
many 
($\sim N$) charges that commute with the Hamiltonian (\ref{Hamiltonian}) and
with each other, whose generating function is the so-called transfer
matrix (\ref{transfer}).

The main motivation for the present work is to address
the problem of
completeness, by which we mean here
whether the Bethe equations have too many, too few, or just the right
number of solutions to describe all the distinct eigenvalues of the
 transfer matrix.
This question is particularly interesting when $q$ is a root of unity,
in which case the Hamiltonian is neither Hermitian nor 
normal, and in fact has Jordan
cells~\cite{Dubail:2010zz,Vasseur:2011fi,MorinDuchesne:2011kd}; and therefore
the number of (ordinary) eigenvectors is less than $2^{N}$ -- the total number of
states.

For the case that $q$ is a root of unity, we
propose a formula for the number of admissible solutions of the Bethe
equations in terms of dimensions \cite{Gainutdinov:2012mr,
Gainutdinov:2012qy} of irreducible representations
of the Temperley-Lieb algebra \cite{Temperley:1971iq}, see Eq.  
(\ref{mainconjecture2}). 
We also propose a formula for the degeneracies of
the transfer matrix eigenvalues in terms of dimensions of tilting
$\qg$-modules, see Eq. (\ref{dimconjecture}) 
These formulas include corrections that appear if two
or more tilting modules are degenerate 
in eigenvalues of the transfer matrix. 
For the XX case ($q=e^{i \pi/2}$), we give explicit formulas for the 
number of admissible solutions and degeneracies, see Eqs. (\ref{p2resultsagain}) 
and (\ref{p2degen}), respectively.
These conjectures, which we have checked up to at least $N=8$, are indeed
consistent with the Bethe ansatz solution for this 
model being complete, 
or Eq. (\ref{completeness}) is satisfied.

An important aspect of these conjectures is the definition of an
admissible solution.  As is the case for the periodic chain (see e.g.
\cite{Nepomechie:2014hma} and references therein), the Bethe equations
(\ref{BAE}) admit singular solutions (i.e., solutions that contain
$\pm \eta/2$).  However, such solutions do not
correspond to eigenvalues and eigenvectors of  
the model (\ref{Hamiltonian}), and
therefore, are not admissible. In the language of
\cite{Nepomechie:2014hma}, all singular solutions of 
the model (\ref{Hamiltonian}) are
``unphysical''; i.e., there are no ``physical'' singular solutions.

Moreover, when $q$ is a root of unity, as is the case for the periodic
XXZ chain \cite{Baxter:1972wg, Fabricius:2000yx, Baxter:2001sx, Tarasov:2003xz}, the
Bethe equations (\ref{BAE}) admit continuous solutions (``algebraic
variety of positive dimension''), in addition to the usual discrete
solutions (``algebraic variety of dimension~$0$'').  However, we
restrict our attention to the latter, which are sufficient
to obtain all the distinct  eigenvalues of the transfer matrix. 
The former are important only
for the construction of the eigenvectors and generalized eigenvectors, 
which we do not discuss here.

The outline of this paper is as follows. In section \ref{sec:XXX}, we
consider the isotropic (XXX) limit $q\rightarrow 1$. In section
\ref{sec:XXZgenericq}, we consider the case of generic values of $q$. 
Our main conjectures are in section \ref{sec:XXZrootunity}, 
where we consider the root of unity case. We briefly
discuss our results in section \ref{sec:discuss}.  Background 
material, special cases and numerical results are provided in the
appendices.  Specifically, the construction of the transfer matrix,
its important properties, and the algebraic Bethe ansatz are reviewed
in appendix \ref{sec:aba}.  The Temperley-Lieb algebra and its
relation to the model (\ref{Hamiltonian}) are briefly reviewed in
appendix \ref{sec:TL}.  The case $p=2$, which can be treated
analytically, is analyzed in appendix \ref{sec:p2}. Examples of cases where 
two or more tilting modules are degenerate are individually analyzed 
in appendix \ref{sec:deviations}. Finally, numerical solutions 
of the Bethe equations up to $N=8$ are displayed in tables in appendix \ref{sec:numerical}.

\section{XXX}\label{sec:XXX}

In the limit $\eta \rightarrow 0$, the Hamiltonian
(\ref{Hamiltonian}) becomes $su(2)$-invariant
\be
H = \sum_{k=1}^{N-1} \vec\sigma_k \cdot \vec\sigma_{k+1}  \,,
\label{HXXX}
\ee
the expression (\ref{energy}) for the eigenvalues becomes\footnote{We
rescale the Bethe roots $\lambda_{j}
\mapsto - i\eta \lambda_{j} $ before taking $\eta \rightarrow 0$.}
\be
E = -2  \sum_{k=1}^{M}
\frac{1}{\lambda_{k}^{2} +\frac{1}{4}} + N-1 \,,
\label{energyXXX}
\ee
and the hyperbolic Bethe equations (\ref{BAE}) become
rational
\be
\left(\lambda_{k} + \frac{i}{2}\right)^{2N}
\prod_{\scriptstyle{j \ne k}\atop \scriptstyle{j=1}}^M
(\lambda_{k} - \lambda_{j} - i)(\lambda_{k} + \lambda_{j} - i)
=\left(\lambda_{k} - \frac{i}{2}\right)^{2N}
\prod_{\scriptstyle{j \ne k}\atop \scriptstyle{j=1}}^M
(\lambda_{k} - \lambda_{j} + i)(\lambda_{k} + \lambda_{j} + i) \,, \non \\
k = 1 \,, 2\,, \ldots \,, M \,, \qquad
M = 0\,, 1\,, \ldots\,,  
\Bigl\lfloor\frac{N}{2}\Bigr\rfloor 
\,. \qquad
\label{BAEXXX}
\ee

The Bethe equations have the reflection symmetry $\lambda_{k} \mapsto
-\lambda_{k}$, while keeping the other $\lambda$'s (i.e. 
$\lambda_{j}$ with $j \ne k$) unchanged.  Moreover, any solution with $\lambda_{k}=0$ must be
discarded, since the corresponding Bethe vector is not an eigenvector
of the Hamiltonian (see e.g. \cite{Fendley:1994cz} and
Appendix \ref{sec:aba}).  Hence, we define a solution $\{
\lambda_{1}\,, \ldots \,, \lambda_{M}\}$ of the Bethe equations
(\ref{BAEXXX}) to be \textit{admissible} if all the $\lambda_{k}$'s are finite
and pairwise distinct (no two are equal), and if each $\lambda_{k}$
satisfies either
\be
\Re e(\lambda_{k} ) > 0   \label{admissibleXXXa}
\ee
or
\be
\Re e(\lambda_{k} ) = 0  \qquad \mbox{ and  } \qquad \Im m(\lambda_{k} ) >
0  \,. \label{admissibleXXXb}
\ee
Note that, according to this definition, singular
solutions $\{ i/2\,, -i/2 \,, \ldots \}$ are
not admissible. As usual, due to the permutation symmetry of the system of 
Bethe equations, the order of the $\lambda$'s in any solution
$\{\lambda_{1} \,, \ldots \,, \lambda_{M}\}$ is irrelevant.

According to the Clebsch-Gordan theorem  
for $su(2)$, the Hilbert space of 
the XXX chain, 
the $N$-fold tensor product of spin-$1/2$ representations 
$V_{\frac{1}{2}}$,  has the decomposition
\be
\bigl(V_{\frac{1}{2}}\bigr)^{\otimes N} = \bigoplus_{j=0(1/2)}^{N/2} d_j 
V_j,
\label{XXXdecomposition}
\ee
where the sum starts from $j=0$ for even $N$ and $j=1/2$ for odd $N$. 
Moreover, $V_{j}$ denotes a spin-$j$ irreducible representation of $su(2)$, 
and the multiplicity $d_{j}$ is given by
\be
d_{j} = {N\choose \frac{N}{2}-j} - {N\choose \frac{N}{2}-j-1} \,,
\qquad \qquad d_{j} = 0 \quad {\rm for  }\quad j > \frac{N}{2} \,.
\label{dj}
\ee

Each admissible solution $\{
\lambda_{1}\,, \ldots \,, \lambda_{M}\}$ corresponds  to a direct summand
$V_{j}$ in the decomposition~\eqref{XXXdecomposition},  with spin $j=\frac{N}{2}-M$. Indeed, as in the case for the periodic XXX chain \cite{Faddeev:1996iy}, the
Bethe states are $su(2)$ highest-weight states and they can be constructed within the algebraic Bethe ansatz, see~\eqref{Bethestate} and
(\ref{highestweight}).
Moreover, we expect that there is a one-to-one correspondence between distinct
admissible solutions $\{ \lambda_{1}\,, \ldots \,, \lambda_{M}\}$ and distinct highest-weight vectors of spin $j=\frac{N}{2}-M$.
Hence, for given values of $N$ and $M$, we conjecture that the 
number ${\cal N}(N,M)$ of admissible solutions of the Bethe equations 
is given by
\be
{\cal N}(N,M) = d_{\frac{N}{2}-M} = {N\choose M} - {N\choose M-1} \,.
\label{conjectureXXX}
\ee
For the periodic XXX chain, it is generally believed that the
number of solutions of the corresponding Bethe equations is
also given by (\ref{conjectureXXX}), see e.g. \cite{Faddeev:1996iy,
Essler:2005}.  However, the situation there is actually more subtle
due to the existence of physical singular solutions \cite{Hao:2013jqa}.

\begin{table}[htb]
    \small 
  \centering
  \begin{tabular}{|l|c|c|c|c|c|c|c|}\hline
    \diagbox[width=2.5em]{$N$}{$M$} & 0 & 1 & 2 & 3\\
    \hline
     2 & 1 & 1 & & \\
     3 & 1 & 2 & & \\
     4 & 1 & 3 & 2 & \\
     5 & 1 & 4 & 5 & \\
     6 & 1 & 5 & 9 & 5 \\
     7 & 1 & 6 & 14 & 14 \\
    \hline
   \end{tabular}
   \caption{\small The number ${\cal N}(N,M)$ of admissible solutions
   of the XXX Bethe equations (\ref{BAEXXX})
   for given values of $N$ and $M$.}
   \label{table:XXX}
\end{table}

Since $\dim  V_{j} = 2j+1$, it is 
also natural to conjecture that the number 
or degeneracy ${\cal D}(N,M)$ of 
eigenvalues of the transfer matrix, 
see its definition in~(\ref{transfer}) and \eqref{RXXX} 
at $q=1$, corresponding to each admissible solution is given by
\be
{\cal D}(N,M) = \dim V_{\frac{N}{2}-M}  = N - 2M +1 \,.
\label{dimconjectureXXX}
\ee
The expressions (\ref{conjectureXXX}) and (\ref{dimconjectureXXX}) 
satisfy the well-known identity
\be
\sum_{M=0}^{\lfloor\frac{N}{2}\rfloor} {\cal N}(N,M)\, {\cal D}(N,M) 
= 2^{N}\,,
\label{completenessXXX}
\ee
signifying the completeness of the solution.

Using homotopy continuation \cite{BHSW1} (see also
\cite{Hao:2013jqa} and references therein for further details),
we have solved  (\ref{BAEXXX}) numerically up to $N=7$. 
 The admissible solutions up to $N=6$ are presented in Table \ref{table:XXXsltns}.
The numbers ${\cal N}(N,M)$ of admissible solutions that we have found are reported in Table~\ref{table:XXX}. 
(For $M=0$, there are no Bethe roots but there is 
nevertheless an eigenvector (\ref{reference}),
so we define ${\cal N}(N,0) = 1$.)
These numbers coincide with the conjectured
values (\ref{conjectureXXX}). As an independent check, 
starting from the transfer matrix~\eqref{transfer}, \eqref{RXXX} at $q=1$
we have explicitly determined each of the transfer matrix eigenvalues 
$\Lambda(u)$ as polynomials in 
$u$ \footnote{Direct diagonalization of the (symbolic) transfer matrix $t(u)$
does not yield the eigenvalues as polynomials in $u$. We instead  
proceed by first finding the (numerical) eigenvectors $|v\rangle$ of the 
(numerical) matrix $t(u_{0})$ for some generic numerical value $u_{0}$. Then, by acting 
with $t(u)$ (whose matrix elements are polynomials in $u$)
on each $|v\rangle$, we read off the corresponding 
eigenvalue $\Lambda(u)$ as a polynomial in $u$. Note that, by virtue 
of the commutativity property (\ref{commutativity}), the eigenvalues do not 
depend on the choice of $u_{0}$.}; 
then, by solving the T-Q equation (\ref{LambdaXXX}) for 
$Q(u)$ and finally finding the zeros of $Q(u)$, we have obtained the 
corresponding Bethe roots. The results match with those obtained by 
directly solving the Bethe equations. The number of eigenvalues 
corresponding to each admissible solution also coincide with 
(\ref{dimconjectureXXX}).

\section{XXZ: generic $q$}\label{sec:XXZgenericq}

We now consider the Bethe equations (\ref{BAE}) for generic values of
$q$, i.e., when $q$ is not a root of unity. These equations have the reflection symmetry $\lambda_{k} \mapsto
-\lambda_{k}$  (while keeping the other $\lambda$'s unchanged), 
and the periodicity $\lambda_{k} \mapsto
\lambda_{k} + i \pi$.  We again exclude $\lambda_{k}=0$, as well as
$\lambda_{k}=\frac{i \pi}{2}$. (See Appendix \ref{sec:aba}.) Hence, we define a solution
$\{ \lambda_{1}\,, \ldots \,, \lambda_{M}\}$ of the Bethe equations (\ref{BAE})
to be \textit{admissible} if all the $\lambda_{k}$'s are finite and pairwise distinct (no
two are equal), and if each $\lambda_{k}$ satisfies
either
\be
\Re e(\lambda_{k} ) > 0  \qquad  \mbox{ and  } \qquad
-\frac{\pi}{2} <  \Im m(\lambda_{k} ) \le \frac{\pi}{2}
\label{admissibleXXZa}
\ee
or
\be
\Re e(\lambda_{k} ) = 0  \qquad \mbox{ and  } \qquad 0 <  \Im m(\lambda_{k} ) < \frac{\pi}{2}   \,.
\label{admissibleXXZb}
\ee

The Hamiltonian (\ref{Hamiltonian}) is invariant under
the quantum group $\qg$, which is the symmetry of the model.
This symmetry is generated by the $S^{\pm}$ and $S^z$ operators that 
now satisfy the quantum-group relations
\be\label{qsl-rel}
[S^z,S^{\pm}]=\pm S^{\pm},\qquad [S^+,S^-] = [2S^z]_{q},
\qquad [x]_{q} \equiv \frac{q^{x}-q^{-x}}{q-q^{-1}}\,, 
\ee
which are just $q$-deformed versions of the usual relations of $su(2)$, or rather $U sl(2)$.
The Bethe vectors (\ref{Bethestate}) are $\qg$ highest-weight states
(\ref{highestweight}), (\ref{Szeig}). For generic values of $q$,
the irreducible representations of $\qg$ are
isomorphic to those of $U sl(2)$. (See e.g. \cite{Chari:1994pz} and references
therein.) The Hilbert space has the same decomposition as in 
the XXX case (\ref{XXXdecomposition}), except that $V_{j}$ is now a spin-$j$ irreducible 
representation of $\qg$ with dimension $2j+1$.
We similarly expect that there is a one-to-one correspondence between distinct
admissible solutions $\{ \lambda_{1}\,, \ldots \,, \lambda_{M}\}$ and 
distinct direct summands isomorphic to $V_{j}$ with $j=\frac{N}{2}-M$.
Hence, we conjecture that the number ${\cal N}(N,M)$
of admissible solutions of the Bethe equations (\ref{BAE}) for generic values of
$q$ is again given by (\ref{conjectureXXX}); and that the number ${\cal D}(N,M)$ of 
eigenvalues of the transfer matrix corresponding 
to each admissible solution is again given by 
(\ref{dimconjectureXXX}).

In order to check these conjectures, it is convenient to rewrite the
Bethe equations (\ref{BAE}) in polynomial form
\be
\left(q x_{k} -1\right)^{2N}
\prod_{\scriptstyle{j \ne k}\atop \scriptstyle{j=1}}^M
(x_{k} - q^{2} x_{j})(x_{k}\, x_{j} - q^{2})
=\left(x_{k} - q\right)^{2N}
\prod_{\scriptstyle{j \ne k}\atop \scriptstyle{j=1}}^M
(q^{2} x_{k} - x_{j}) (q^{2} x_{k}\, x_{j} -1) \,, \non \\
k = 1 \,, 2\,, \ldots \,, M \,, \qquad
M = 0\,, 1\,, \ldots\,,  
\Bigl\lfloor\frac{N}{2}\Bigr\rfloor  \,,  \qquad\qquad\qquad
\label{BAEx}
\ee
where 
\begin{equation}
x_{k}=e^{2\lambda_{k}}\,.
\end{equation}
 For admissible solutions, each $x_{k}$ satisfies
either
\be
|x_{k}| > 1
\label{admissibleXXZxa}
\ee
or
\be
|x_{k}| = 1  \qquad \mbox{ and  } \qquad 0 < \arg (x_{k} ) < \pi   \,.
\label{admissibleXXZxb}
\ee
We have solved this system numerically with $\eta=0.1$ up to $N=7$.
 The 
admissible solutions up to $N=6$ are presented in Table \ref{table:XXZsltns}.
The numbers ${\cal N}(N,M)$ of admissible solutions that we have 
found are reported in Table
\ref{table:XXZqgen}. These results are the same as for the XXX case,
and therefore coincide with the conjectured values (\ref{conjectureXXX}). 
 We have also confirmed that the degeneracy of the eigenvalues 
is again given by (\ref{dimconjectureXXX}).
We have obtained similar results for 
$\eta=i/2$, in which case $|q|=1$ (and therefore the Hamiltonian is 
critical; although the Hamiltonian is not Hermitian or even normal, 
it is nevertheless diagonalizable) but $q$ is not a root of unity. 

\begin{table}[htb]
    \small
  \centering
  \begin{tabular}{|l|c|c|c|c|c|c|c|}\hline
    \diagbox[width=2.5em]{$N$}{$M$} & 0 & 1 & 2 & 3\\
    \hline
     2 & 1 & 1 & & \\
     3 & 1 & 2 & & \\
     4 & 1 & 3 & 2 & \\
     5 & 1 & 4 & 5 & \\
     6 & 1 & 5 & 9 & 5 \\
     7 & 1 & 6 & 14 & 14 \\
    \hline
   \end{tabular}
   \caption{\small The number ${\cal N}(N,M)$ of admissible solutions
   of the Bethe equations (\ref{BAE}), (\ref{BAEx})
   for given values of $N$ and $M$ and $\eta=0.1$ (a generic value).}
   \label{table:XXZqgen}
\end{table}

In view of the algebraic Bethe ansatz construction for the
eigenstates~\eqref{Bethestate}, our conjectures say that distinct
Bethe states correspond to distinct admissible solutions of the Bethe
equations.
Moreover, we also assume that, to each eigenvalue of the transfer matrix,
there corresponds a unique admissible solution.  Indeed, in the case
of generic $q$, for a given eigenvalue $\Lambda(u)$, we expect that
the T-Q equation (\ref{Lambda}) has a unique (up to rescaling)
solution $Q(u)$, which implies a corresponding unique admissible
solution $\{\lambda_k\}$. (We have checked this numerically for 
small values of $N$. Indeed, as in the XXX case, the Bethe roots 
obtained in this way match with those obtained by directly solving 
the Bethe equations.)
If this is true, that would imply that the spectrum of
the transfer matrix on the $\qg$ highest-weight states is
non-degenerate. 
(For the 
periodic XXX chain, it has been 
shown that the spectrum of the transfer matrix 
on $sl(2)$ highest-weight states is non-degenerate~\cite{Mukhin:2009}.)

\section{XXZ: $q=e^{i \pi/p}$}\label{sec:XXZrootunity}

We now consider the Bethe equations (\ref{BAE}) when $q$ is a
primitive $2p^{th}$ root of unity: $q=e^{i \pi/p}$, where 
$p=2, 3, 4, \ldots$
Inspection of (\ref{BAEx}) shows that, for such cases, the
top degree terms can cancel, suggesting that the system is
qualitatively different from the generic $q$ case.

A particularly interesting new feature is that the Bethe equations now
admit continuous solutions, in addition to the usual discrete
solutions. For example, the following set of $p$ elements
\be
\{ \lambda_{0} + \frac{i \pi}{2p}(p-1) \,, \lambda_{0} + \frac{i
\pi}{2p}(p-3)\,, \ldots \,, \lambda_{0} - \frac{i
\pi}{2p}(p-3)\,, \lambda_{0} - \frac{i \pi}{2p}(p-1) \}
\label{exactcompleterstring}
\ee
is an exact solution of the Bethe equations (\ref{BAE}) with $\eta= i \pi/p$
and $M=p$, for arbitrary values of $\lambda_{0}$. Such solutions have
been discussed in the context of periodic chains \cite{Baxter:1972wg,
Fabricius:2000yx, Baxter:2001sx, Tarasov:2003xz}, and are called
``exact complete $p$-strings.'' In the parlance of algebraic geometry, such solutions have positive dimension. In
contrast, the usual discrete solutions instead have dimension $0$.
The solutions (\ref{exactcompleterstring}) are related to 
certain degeneracies of the model: 
the corresponding energy (as well as eigenvalue $\Lambda(u)$ obtained 
from the T-Q equation (\ref{Lambda})) 
is the same as for the reference (pseudovacuum) state. 
Bethe states corresponding to 
such solutions are \textit{prima facie} null; a regularization scheme and a suitable 
limiting procedure are needed to obtain non-null states (see 
\cite{Tarasov:2003xz} and references therein for the periodic case).

\subsection{Admissible solutions}
We restrict our attention here to the usual discrete solutions, which
are sufficient to obtain all the distinct eigenvalues of
the transfer matrix.\footnote{The transfer matrix and the 
Hamiltonian generally have the same number of distinct eigenvalues. 
However, there are exceptions, such as the case $p=4$ and $N=8$, where 
the number of distinct eigenvalues is 43 and 41 for the transfer matrix and  
Hamiltonian, respectively. In other words, for this case there are 43 
admissible solutions: 39 solutions give (through Eq.(\ref{energy})) 39 
distinct energies, while 2 pairs of solutions give equal values of the 
energy, for a total of only 41 distinct energies. Another 
exception is the case $p=2$ and $N=9$, where 
the number of distinct eigenvalues is 81 and 57 for the transfer matrix and  
Hamiltonian, respectively. We expect that such ``mismatches'' occur for 
other values of $p$ and $N$, but we have not made an effort to study them 
systematically.} 
Indeed, the union $s_1 \cup s_2$ of a discrete solution
$s_1$ and an exact complete $p$-string solution $s_2$ is again a
solution; hence, adding a $p$-string does not change the
eigenvalue corresponding to the initial discrete solution.

We therefore define an admissible solution of the Bethe equations as
before in (\ref{admissibleXXZa}) and (\ref{admissibleXXZb}), except with the
additional requirement that the solution should {\em not} contain
the exact complete $p$-string (\ref{exactcompleterstring}). 

\subsection{Generalized eigenvalues and tilting modules}
As already noted in the Introduction,  
non-trivial Jordan-block structure for $H$ appears at 
roots of unity. Therefore, we now consider {\em generalized} eigenvalues of 
the transfer matrix (and of the Hamiltonian); i.e., eigenvalues $\Lambda(u)$ corresponding to 
generalized eigenvectors $|v\rangle$ that are defined as (also called root vectors)
\be
\bigl(t(u)-\Lambda(u)\mathbf{1} \bigr)^2 |v\rangle=0 \,,
\label{geneig}
\ee
or equivalently 
\be
t(u)\, |v\rangle=\Lambda(u)\, |v\rangle + |v'\rangle  \quad \mbox{    and   
} \quad t(u)\, |v'\rangle=\Lambda(u)\, |v'\rangle \,.
\label{geneig-2}
\ee
The power in (\ref{geneig}) is $2$ because there are Jordan cells of 
maximum rank $2$, and here $|v\rangle$ and $|v'\rangle$ belong to a Jordan cell of 
rank $2$.\footnote{The function {\tt 
Eigenvalues[]} in Mathematica computes generalized eigenvalues.} 
So, we have  the
number of eigenvectors less than $2^N$ but the number of generalized eigenvectors
is exactly $2^N$.

For $q=e^{i \pi/p}$, the $N$-fold tensor product of spin-1/2
representations decomposes into a direct sum of certain indecomposable modules $T_j$ of
$\qg$ characterized by spin $j$. More precisely, these direct 
summands $T_j$ are so-called \textit{tilting} $\qg$-modules which are
(i)  composed of the standard spin modules and (ii) satisfy a
self-duality condition or invariance under the
adjoint~$\cdot^{\dagger}$ operation   (see~\cite{Gainutdinov:2012ms}
for a short review in the context of open spin chains.)  These two
properties usually lead to a complicated structure of indecomposable
but reducible modules, i.e., those having invariant subspaces but
cannot be split onto a direct sum.  The structure of the tilting
$\qg$-modules was studied in
many works~\cite{Pasquier:1989kd,Martin:1991pk,Chari:1994pz,Read:2007qq,
Gainutdinov:2012mr} and in brief it is the following:
if $2j+1$ is bigger than $p$ and not $0$ modulo $p$ then each $T_j$
is composed of the spin-$j$ 
(or $V_j$ in our notations) and the spin-$(j-s(j))$ modules,
where\footnote{$(j\ {\rm mod} \ p)$ is the remainder on division of
$j$ by $p$.} $s(j)=(2j+1)\;{\rm mod}\;p$, such that the former is a submodule; otherwise, $T_j$ is irreducible. So, in particular we have the dimensions
\be
\dim T_j =
\begin{cases}
2j+1,&\qquad 2j+1\leq p \quad\text{or}\quad s(j)=0,\\
2(2j+1-s(j)),&\qquad \text{otherwise}.
\end{cases}
\label{dimTj}
\ee

Equipped with this information about $T_j$'s we can write a 
decomposition of the 
XXZ spin-$\frac{1}{2}$ chain as
\be
\bigl(V_{\frac{1}{2}}\bigr)^{\otimes N} = \bigoplus_{j=0(1/2)}^{N/2} d^0_j T_j,
\label{decomposition}
\ee
where the sum starts from $j=0$ for even $N$ and $j=1/2$ for odd $N$.
The important point is that the multiplicities $d^0_j$ of these $T_j$
modules can be explicitly computed using representation
theory~\cite{Gainutdinov:2012mr} and are given by
the dimensions $d_{j}^{0}$ of irreducible representations of the
Temperley-Lieb (TL) algebra with the fugacity or loop parameter
$\delta=2\cos{\frac{\pi}{p}}$, 
see Appendix~\ref{sec:TL}
for definitions:
\be
d_{j}^{0} = \sum_{n\ge 0} d_{j+ n p}
- \sum_{n \ge t(j)+1} d_{j+ n p -1 -2(j
\, {\rm  mod } \, p)} \,, \qquad (j\ {\rm mod} \ p) \ne p-\frac{1}{2}
\,, \frac{p-1}{2} \,,
\label{dj0}
\ee
where $d_{j}$ is given by (\ref{dj}), and
\be
t(j) = \left\{ \begin{array}{ll}
1 & \ {\rm for  }\  (j\ {\rm mod} \ p) > \frac{p-1}{2} \,, \\
0 & \ {\rm for  }\  (j\ {\rm mod} \ p) < \frac{p-1}{2} \,.
\end{array} \right.
\label{tj}
\ee
If $(j\ {\rm mod} \ p) = p-\frac{1}{2} \,, \frac{p-1}{2}$, then
$d_{j}^{0} =d_{j}$. Note that one can check
\be
\sum_{j=0(1/2)}^{N/2}d^0_j \dim T_j = 2^N.
\ee

Since the transfer matrix commutes with the generators of 
$\qg$, see~(\ref{tqgsym}), all the (generalized) eigenvectors~\eqref{geneig} in a given (direct summand isomorphic to the) tilting module 
$T_{j}$ have the same (generalized) eigenvalue of the transfer matrix. 
It is the indecomposable but reducible tilting modules that are
responsible for the Jordan cells structure in the Hamiltonian and the
presence of the generalized eigenvectors~$|v\rangle$: they live in
heads of the tilting modules while their partners $ |v'\rangle$,
see~\eqref{geneig-2}, live in the socle -- the irreducible submodule
of~$T_j$.

\subsection{Main conjectures}\label{sec:conj}
Assuming that there is at most one admissible solution of the Bethe 
equations 
for the generalized eigenvalue in each direct summand isomorphic to the $\qg$-module $T_j$, 
the number ${\cal N}(N,M)$ of admissible solutions
of the Bethe equations (\ref{BAE}) with $\eta= i \pi/p$ satisfies the 
inequality 
\be
{\cal N}(N,M) \le d_{\frac{N}{2}-M}^{0} \,,
\label{mainconjecture}
\ee
where $d_{j}^{0}$ is given by (\ref{dj0})-(\ref{tj}), and we 
have used the relation $j=\frac{N}{2}-M$ stated in~(\ref{spinj}). We argue in Appendix 
\ref{sec:deviations} that 
${\cal N}(N,M) < d^{0}_{\frac{N}{2}-M}$ 
 when two or more tilting modules become 
degenerate 
in the sense that the generalized eigenvalues of the transfer 
matrix corresponding to direct
summands $T_j$ and $T_k$ in~\eqref{decomposition}, for distinct $j$ and $k$, are equal.  This
suggests that the conjecture can be sharpened to the following:
\be
{\cal N}(N,M) = d_{\frac{N}{2}-M}^{0} - n_{\frac{N}{2}-M}\,,
\label{mainconjecture2}
\ee
where $n_j$ is the number of direct summands $T_j$ that are degenerate with other 
tilting modules $T_{k}$ with $k >j$ in the decomposition (\ref{decomposition}). 
We note that exact complete $p$-string 
solutions (\ref{exactcompleterstring}) are needed to construct the 
Bethe states corresponding to such degenerate tilting modules.

 We can similarly conjecture that the number or
degeneracy ${\cal D}(N,M)$ of the 
generalized eigenvalues of the
transfer matrix corresponding to each admissible solution satisfies
the inequality\footnote{We note that the numbers ${\cal N}(N,M)$ and ${\cal D}(N,M)$ depend also on $p$, as the dimensions of irreducible TL representations and of tilting modules do, but we do not use this dependence in notations for brevity.}
\be
{\cal D}(N,M) \ge \dim T_{\frac{N}{2}-M} \,,
\label{dimconjectureA}
\ee
where $\dim T_{j}$ is given by (\ref{dimTj}). 
We can also sharpen this conjecture by introducing $n_{j k}$, which
we define as the number of tilting modules $T_{k}$ (with $k < j$) 
in the decomposition (\ref{decomposition})
that are degenerate with~$T_{j}$.\footnote{If $d_{j}^{0} > 1$ (i.e., 
there is more than one copy of $T_j$) and $n_{j k}$ is nonzero for 
some $k<j$, then it is implicit that {\em each} copy of $T_j$ is degenerate with 
$n_{j k}$ copies of $T_k$. This assumption appears to be satisfied in all the examples that we have 
considered.} 
(We define $n_{j k} = 0$ for $k \ge j$.)
Then, we conjecture that the
degeneracy of an eigenvalue of the transfer matrix 
(corresponding to a given admissible solution $\{\lambda_{1}, \ldots , 
\lambda_{M}\}$) equals\footnote{The sum in 
(\ref{dimconjecture}) is over all $k<j$ in the decomposition 
(\ref{decomposition}); hence, it starts from $k=0$ for even $N$ and 
$k=1/2$ for odd $N$. Strictly speaking, 
the restriction $k<j$ is not necessary 
since $n_{j k} = 0$ for $k \ge j$, but in this way we
emphasize the relevant contributions.}
\be
{\cal D}(N,M) 
= \dim T_{j} + \sum_{k<j} n_{j k}\, \dim T_{k} 
\, , \quad \text{with}\;\; j=\ffrac{N}{2}-M\,.
\label{dimconjecture}
\ee
It is not obvious that the degeneracy ${\cal D}(N,M)$ is the same for all admissible
solutions with a given value of $M$ (as it is in the generic case), but it is
so for the cases that we have considered.  We therefore further conjecture that the numbers
${\cal D}(N,M)$ (and also $n_{jk}$) do not actually depend on a
particular solution $\{\lambda_{1}, \ldots , \lambda_{M}\}$.

\begin{table}[htb]
    \small
   \centering
   \subfloat[$q=e^{i \pi/2}$]
  {\begin{tabular}{|l|c|c|c|c|c|c|c|}\hline
    \diagbox[width=2.5em]{$N$}{$M$}  & 0 & 1 & 2 & 3 & 4\\
    \hline
     2 & 1 & 0 & & & \\
     3 & 1 & 2 & & & \\
     4 & 1 & 2 & 0 & & \\
     5 & 1 & 4 & 4 [5] & & \\
     6 & 1 & 4 & 4 [5] & 0 & \\
     7 & 1 & 6 & 12 [14] & 8 [14] & \\
     8 & 1 & 6 & 12 [14] & 8 [14] & 0 \\
     9 & 1 & 8 & 24 [27] & 32 [48] & 16 [42] \\
    \hline
   \end{tabular}}\qquad
  \subfloat[$q=e^{i \pi/3}$]
  {\begin{tabular}{|l|c|c|c|c|c|c|c|}\hline
    \diagbox[width=2.5em]{$N$}{$M$} & 0 & 1 & 2 & 3 & 4\\
    \hline
     2 & 1 & 1 & & & \\
     3 & 1 & 1 & & & \\
     4 & 1 & 3 & 1 &  & \\
     5 & 1 & 4 & 1 &  & \\
     6 & 1 & 4 & 9 & 1  & \\
     7 & 1 & 6 & 13 & 1 & \\
     8 & 1 & 7 & 13 & 27 [28] & 1 \\
    \hline
   \end{tabular}}\qquad
   \subfloat[$q=e^{i \pi/4}$]
   {\begin{tabular}{|l|c|c|c|c|c|c|c|}\hline
    \diagbox[width=2.5em]{$N$}{$M$} & 0 & 1 & 2 & 3 & 4\\
    \hline
     2 & 1 & 1 & &  & \\
     3 & 1 & 2 & &  & \\
     4 & 1 & 2 & 2 &  & \\
     5 & 1 & 4 & 4 &  & \\
     6 & 1 & 5 & 4 & 4  & \\
     7 & 1 & 6 & 14 & 8  & \\
     8 & 1 & 6 & 20 & 8 & 8\\
    \hline
   \end{tabular}}\qquad
   \subfloat[$q=e^{i \pi/5}$]
   {\begin{tabular}{|l|c|c|c|c|c|c|c|}\hline
   \diagbox[width=2.5em]{$N$}{$M$} & 0 & 1 & 2 & 3 & 4\\
    \hline
     2 & 1 & 1  & &  & \\
     3 & 1 & 2  & &  & \\
     4 & 1 & 3  & 2  &  & \\
     5 & 1 & 3  & 5  &  & \\
     6 & 1 & 5  & 8  & 5  & \\
     7 & 1 & 6  & 8  & 13  & \\
     8 & 1 & 7  & 20 & 21  & 13\\
    \hline
   \end{tabular}}
   \caption{\small The number ${\cal N}(N,M)$ of admissible solutions
   of the Bethe equations (\ref{BAE}), (\ref{BAEx})
   for given values of $N$, $M$ and $q$. Numbers within brackets 
   are the values of $d_{\frac{N}{2}-M}^{0}$ (\ref{dj0}), when different 
   from~${\cal N}(N,M)$.}
   \label{table:XXZp}
\end{table}

The two sets of integers $\{ n_{j} \}$ and $\{ n_{jk} \}$ should be related by
\be
n_{j} &=& \sum_{m\geq0}(-1)^{m}\sum_{j_{0}, j_{1}, \ldots, j_{m}=0 
(1/2)}^{N/2} 
d^{0}_{j_{m}}\, n_{j_{m} j_{m-1}}\, n_{j_{m-1} j_{m-2}} \cdots  
n_{j_{0} j} \,, \non \\
&=&  \sum_{j_{0}=0 (1/2)}^{N/2} d^{0}_{j_{0}}\, n_{j_{0} j} - 
\sum_{j_{0}, j_{1}=0 (1/2)}^{N/2} d^{0}_{j_{1}}\,  n_{j_{1} j_{0}}\, n_{j_{0} j} + \ldots
\label{njnjk}
\ee
The idea is that, if no more than two 
(non-isomorphic) tilting modules 
are degenerate, then only the $m=0$ term in (\ref{njnjk}) is nonzero; 
however, if 3 tilting modules are degenerate (e.g. the case $p=2$, 
$N=9$, for which the modules $T_{\frac{9}{2}}$, $T_{\frac{5}{2}}$ and
$T_{\frac{1}{2}}$ are degenerate, see (\ref{njkp2n9})), then the $m=1$ term in 
(\ref{njnjk}) provides a nonzero correction, etc. 
Indeed, one can verify that the sum rule
\be
\sum_{M=0}^{\lfloor\frac{N}{2}\rfloor} {\cal N}(N,M)\, {\cal D}(N,M) = 2^{N}
\label{completeness}
\ee
is satisfied using (\ref{mainconjecture2}) for 
${\cal N}(N,M)$, (\ref{dimconjecture}) for ${\cal D}(N,M)$, 
and the expression (\ref{njnjk}) for $n_{j}$, 
with {\em arbitrary} $n_{j k}$, except 
that $n_{j k} = 0$ for $k \ge j$ (already noted above), 
and also that $n_{j k}=0$ if $(j-k)\ {\rm mod}\ p \ne 0$, which is 
discussed further below.

For the case $p=2$, we have more explicit results. The number of 
admissible solutions ${\cal N}(N,M)$ for general values of $N$ and $M$ is 
given by
\be
{\cal N}(N,M) =  
 \left\{ \begin{array}{cc}
                       \frac{(N-2)!!}{M! (N-2-2M)!!} & \quad N = {\rm even} \\
		       \frac{(N-1)!!}{M! (N-1-2M)!!} & \quad N = {\rm odd} 
			\end{array} \right.\,,
			\label{p2resultsagain}
\ee 
as shown in Appendix \ref{sec:p2}. We conjecture that the 
degeneracies ${\cal D}(N,M)$ for general values of $N$ and $M$ are 
given by\footnote{In terms of the spin $j=\frac{N}{2}-M$, the 
degeneracies are given by 
\be
{\cal D}_{j} \equiv  {\cal D}(N,\frac{N}{2}-j) = 
\begin{cases}
    2^{\lfloor j\rfloor+1}\,,&\qquad j>0\,, \\
    0\,, &\qquad j=0\,,
\end{cases} \non
\label{p2degen}
\ee
which evidently do not depend on $N$.}
\be
{\cal D}(N,M) = 
\begin{cases}
    2^{\lfloor\frac{N}{2}\rfloor-M+1}\,,&\qquad M < \frac{N}{2}\,, \\
    0\,, &\qquad M = \frac{N}{2} \quad \text{and} \quad N = \text{even} \,.
\end{cases}
\label{p2DNM}
\ee
Indeed, this formula reproduces the results in Table 
\ref{table:XXZpmore} (a) below; and, together with 
(\ref{p2resultsagain}) for 
${\cal N}(N,M)$, satisfies the sum rule (\ref{completeness}). 
Moreover, we propose that the integers $n_{jk}$ in (\ref{dimconjecture}) 
are given, for $p=2$ and $j$ and $k$ integers, by
\be
n_{jk}=
\begin{cases}
    {j-1\choose \frac{1}{2}(j-k)}\frac{2k}{j+k}\,,&\qquad j > k \quad \text{and} \quad 
    (j-k)\ {\rm mod}\ 2 = 0\,, \\
    0\,,&\qquad j \le k\,, \quad \text{or} \quad (j-k)\ {\rm mod}\ 2 \ne 
    0\,, \quad \text{or} \quad k=0 \,, 
\end{cases}
\label{p2njk}
\ee
which do not depend on $N$. We note, as a curiosity, that
$n_{2j,2}$ for $j>1$ is equal to the $j^{th}$ Catalan 
number. For $j$ and $k$ half-odd integers, $n_{jk}=n_{j+\frac{1}{2}, 
k+\frac{1}{2}}$. Indeed, these formulas reproduce all the values of 
$n_{jk}$ for $p=2$ found in Appendix \ref{sec:deviations}, and 
satisfy (\ref{mainconjecture2}) (with ${\cal N}(N,M)$ and
$n_{j}$ given by (\ref{p2resultsagain}) and (\ref{njnjk}) , respectively))
as well as (\ref{dimconjecture}) (with ${\cal D}(N,M)$ given by 
(\ref{p2DNM})).

\begin{table}[htb]
    \small
   \centering
   \subfloat[$q=e^{i \pi/2}$]
  {\begin{tabular}{|l|c|c|c|c|c|c|c|}\hline
    \diagbox[width=2.5em]{$N$}{$M$}  & 0 & 1 & 2 & 3 & 4\\
    \hline
     2 & 4 & 0 & & &  \\
     3 & 4 & 2 & & &  \\
     4 & 8 & 4 & 0 & &  \\
     5 & 8 [6] & 4 & 2 & &   \\
     6 & 16 [12] & 8 & 4 & 0 &  \\
     7 & 16 [8] & 8 [6] & 4 & 2 &  \\
     8 & 32 [16] & 16 [12] & 8 & 4 & 0  \\
     9 & 32 [10] & 16 [8] & 8 [6] & 4 & 2 \\
    \hline
   \end{tabular}}\qquad
   \subfloat[$q=e^{i \pi/3}$]
   {\begin{tabular}{|l|c|c|c|c|c|}\hline
     \diagbox[width=2.5em]{$N$}{$M$} & 0 & 1 & 2 & 3 & 4\\
     \hline
      2 & 3 & 1 & & & \\
      3 & 6 & 2 & & & \\
      4 & 6 & 3 & 1 &  & \\
      5 & 6  & 6 & 2 &  & \\
      6 & 12 & 6 & 3 & 1  & \\
      7 & 12 & 6 & 6 & 2 & \\
      8 & 12 [9] & 12 & 6 & 3  & 1 \\
     \hline
    \end{tabular}}\qquad
    \subfloat[$q=e^{i \pi/4}$]
    {\begin{tabular}{|l|c|c|c|c|c|}\hline
     \diagbox[width=2.5em]{$N$}{$M$} & 0 & 1 & 2 & 3 & 4\\
     \hline
      2 & 3 & 1 & &  & \\
      3 & 4 & 2 & &  & \\
      4 & 8 & 3 & 1 &  & \\
      5 & 8 & 4 & 2 &  & \\
      6 & 8 & 8 & 3 & 1  & \\
      7 & 8 & 8 & 4 & 2 & \\
      8 & 16 & 8 & 8 & 3 & 1 \\
     \hline
    \end{tabular}}\qquad
    \subfloat[$q=e^{i \pi/5}$]
    {\begin{tabular}{|l|c|c|c|c|c|}\hline
    \diagbox[width=2.5em]{$N$}{$M$}& 0 & 1 & 2 & 3 & 4\\
     \hline
      2 & 3  & 1  & &  & \\
      3 & 4  & 2  & &  & \\
      4 & 5  & 3  & 1 &  & \\
      5 & 10  & 4  & 2  &  & \\
      6 & 10  & 5  & 3  & 1  & \\
      7 & 10  & 10  & 4  & 2  & \\
      8 & 10  & 10  & 5 & 3  & 1\\
     \hline
    \end{tabular}}
\caption{\small The number ${\cal D}(N,M)$ of eigenvalues of the transfer 
matrix corresponding to each admissible solution of the Bethe equations (\ref{BAE}), (\ref{BAEx})
for given values of $N$, $M$ and $q$. Numbers within brackets 
are the values of $\dim T_{\frac{N}{2}-M}$ (\ref{dimTj}), when different 
from  ${\cal D}(N,M)$.} 
\label{table:XXZpmore}
\end{table}

The appearance of the extra degeneracies among different tilting 
modules at roots of unity  is not surprising, as we have an extra 
symmetry for the whole family of integrable Hamiltonians. For the 
case $p=2$, this extra symmetry was identified in~\cite[Sec.~2.6.2
and~5]{Gainutdinov:2012qy} with the zero modes of the so-called
lattice W-algebra.  These modes $W^{\pm,r}_0$, with
$r,s\in2\mathbb{N}_0$, are particular operators that commute with $H$
and change the total spin $S^z$ by $\pm2$ and mix the distinct tilting
$\qg$-modules.  These operators satisfy relations
\begin{align}
[W^{+,r}_0,W^{-,s}_0]&=4W^{0,r+s+2}_0-4W^{0,r+s}_0,\label{W-1}\\
 [W^{0,r}_0,W^{+,s}_0]&=-8W^{+,r+s+2}_0+8W^{+,r+s}_0,\\
[W^{0,r}_0,W^{-,s}_0]&=8W^{-,r+s+2}_0-8W^{-,r+s}_0,\label{W-3}
\end{align}
where $W^{0,r}_0$ are spinless zero modes of the W-algebra.  The
relations resemble the loop $sl(2)$ algebra relations and the algebra
of the zero modes $W^{\alpha,r}_0$ was indeed identified with a
subalgebra in it~\cite{Gainutdinov:2012qy}.  For higher roots of
unity, there should exist a similar construction
of the zero modes of the lattice W-algebra, defined 
in~\cite{Gainutdinov:2012qy} for all $p$, and these operators do not 
commute with $S^z$ but do commute with the Cartan $\qg$ generator $K=q^{2S^z}$. So, we might expect a mixing of tilting modules in  sectors  by $S^z$ equal modulo $p$.

We have solved 
the Bethe equations (\ref{BAEx}) with $q=e^{i \pi/p}$ 
numerically for $p = 3, 4, 5$ up to $N=8$, 
 see Tables \ref{table:XXZp3sltns1}-\ref{table:XXZp5sltns2}.
The numbers ${\cal N}(N,M)$ of admissible solutions that we have
found are reported in Table \ref{table:XXZp}.\footnote{The results for $p=2$ were obtained 
using Eq. (\ref{p2results}). The results for $p \ge 3$ with $N=8$ and $M=4$ were 
obtained only by solving the T-Q equations.}
These values are consistent with the conjecture
(\ref{mainconjecture}).  Note that ${\cal N}(N,M)$ is equal to the dimension
$d_{\frac{N}{2}-M}^{0}$ of  the TL irreducible representation for most of the values of $N$ and $M$ that we
have considered.  For the few cases that ${\cal N}(N,M) <
d_{\frac{N}{2}-M}^{0}$, the values of $d_{\frac{N}{2}-M}^{0}$ appear
in the tables within brackets.  We analyze these cases individually in
Appendix \ref{sec:deviations}, and we argue that they are consistent
with the sharpened conjecture (\ref{mainconjecture2}).

The numbers ${\cal D}(N,M)$ of eigenvalues of the transfer matrix 
(\ref{transfer}) corresponding to each admissible solution of the Bethe equations are
reported in Table \ref{table:XXZpmore}.  These values are consistent with the conjecture
(\ref{dimconjectureA}). 
Note that ${\cal D}(N,M)$ is  equal to $\dim T_{\frac{N}{2}-M}$ for most of the values of $N$ and $M$ that we
have considered.  For the few cases that ${\cal D}(N,M) > \dim T_{\frac{N}{2}-M}$, the values of 
$\dim T_{\frac{N}{2}-M}$ appear in the tables within brackets. We also analyze these cases individually in
Appendix \ref{sec:deviations}, and we argue that they are consistent 
with the conjecture (\ref{dimconjecture}).

\section{Discussion}\label{sec:discuss}

 We have proposed formulas (\ref{conjectureXXX}),
(\ref{mainconjecture2}), (\ref{p2resultsagain}) for the number of admissible solutions of the
Bethe equations (\ref{BAE}), as well as formulas
(\ref{dimconjectureXXX}), (\ref{dimconjecture}), (\ref{p2DNM}) for the degeneracies
of the transfer matrix eigenvalues,  
including the root of unity cases $q=e^{i\pi/p}$ with $p\geq2$.  
These formulas are
consistent with the completeness of the solution (\ref{completenessXXX}),
(\ref{completeness}).  We have checked these conjectures up to at
least $N=8$.  We emphasize that we consider here {\em all} the
(admissible) solutions of the Bethe equations, not just those
corresponding to ``good'' states \cite{Pasquier:1989kd,
Juttner:1994sk}.  
The  construction of all the Bethe states remains to be clarified.
Work on this and related questions is now in progress.

We have observed at $p=2$ and $p=3$ large  degeneracies
(in the spectrum of the transfer-matrix)
that cannot be explained just using the representation theory of the
Temperley-Lieb algebra or $\qg$ at roots of unity. 
We expect actually similar degeneracies for all integer $p\geq2$ starting with sufficiently
large $N$, for example, $p=4$ and $N \ge 10$.   Such
degeneracies appear due to a very fine phenomena.  It is similar to
the periodic case where, at roots of unity, there is a much bigger
symmetry of $H$ -- the loop $sl(2)$ algebra (at least for $p=2$
\cite{Deguchi:1999rq}).  This symmetry, additionally to the quantum
group generators, mixes $H$-eigenvectors in sectors modulo $p$. 
We expect a similar phenomena in the boundary case, and for $p=2$ we do have such an extra symmetry written explicitly in terms of $W^{\pm,r}_0$ operators satisfying~\eqref{W-1}-\eqref{W-3}, see the discussion in Sec.~\ref{sec:conj}. 
For all integer values of $p\geq2$, we expect that this extra symmetry commutes with the Cartan operator $K=q^{2S^z}$ (and not with $S^z$). In particular,
tilting modules $T_j$ and $T_k$ might be degenerate only if $|j-k|=0\mod p$.
However, instead of the loop $sl(2)$ symmetry that appears in the periodic case, the extra symmetry in the open case should be a 
subalgebra in the loop $sl(2)$. This is expected to be in 
analogy with the $q$-Onsager approach~\cite{Baseilhac:2005tk} to the open XXZ spin-chain with  diagonal boundary conditions~\cite{Baseilhac:2012qi},
where the generating-spectrum algebra for the finite open chain -- the $q$-Onsager algebra -- is 
a  (co-ideal) subalgebra in the generating-spectrum algebra of the closed/periodic chain,
which is the affine quantum algebra $U_q \widehat{sl}(2)$.

This work raises several interesting questions. Assuming that our
conjectures are correct, it would be interesting to find proofs
and explore more the  role of the
lattice W-algebra symmetry~\cite{Gainutdinov:2012qy} in our context of
open chains 
that may be responsible for the degeneracies of the tilting modules, which 
could help to determine the values of $n_{jk}$ in 
(\ref{dimconjecture}) for 
$p>2$. (For $p=2$, see (\ref{p2njk}).)
It would also be interesting to perform a similar analysis of related
models, such as 
the quantum group invariant XXZ chain with higher spin, and
the periodic XXZ chain.  

In our view, it is remarkable that a system of polynomial equations
can ``know'' so much representation theory.  It is evidence that
Bethe ansatz provides deep links between algebraic geometry, representation
theory and quantum mechanics.

\section*{Acknowledgments}
This paper is dedicated to Rodney J. Baxter on the occasion of his 75th birthday.
We thank Dhagash Mehta for his collaboration at an early
stage of this project.  
AMG thanks Pascal Baseilhac, Hubert Saleur, and Ilya Tipunin for helpful discussions. AMG\ was supported by Humboldt fellowship and RFBR-grant 13-01-00386.
AMG\ wishes also to thank the IPhT in Saclay, LMPT in Tours and Max-Planck Institute in Bonn
for hospitality during the work on this project.
WH's research has been supported by the
Mathematical Biosciences Institute and the National Science Foundation
under Grant DMS 0931642.  The work of RN was supported in part by the
National Science Foundation under Grant PHY-1212337, and by a Cooper
fellowship.  The work of AS was supported by NSF ACI 1440607.

\appendix

\section{Transfer matrix and algebraic Bethe ansatz}\label{sec:aba}

We briefly review here the transfer matrix and algebraic Bethe ansatz
for the model (\ref{Hamiltonian}).  These results were first obtained for a
more general model by Sklyanin \cite{Sklyanin:1988yz}.  The transfer
matrix $t(u)$ is given by
\be
t(u) = \tr_{a} K^{+}_{a}(u)\, T_{a}(u)\, K^{-}_{a}(u)\, \hat T_{a}(u) 
\,,
\label{transfer}
\ee
where $T_{a}(u)$ and $\hat T_{a}(u)$ are the monodromy matrices
\be
T_{a}(u) = R_{a1}(u) \cdots R_{aN}(u)\,, \qquad
\hat T_{a}(u) = R_{aN}(u) \cdots R_{a1}(u) \,,
\ee
the R-matrix is given by
\be
R(u) = \left(\begin{array}{cccc}
\sinh(u+\eta) & 0 & 0 & 0\\
0 & \sinh(u) & \sinh(\eta) & 0 \\
0 & \sinh(\eta) & \sinh(u) & 0 \\
0 & 0 & 0 & \sinh(u+\eta)
\end{array} \right) \,, 
\ee
and the left and right K-matrices are given by the diagonal matrices
\be
K^{+}(u)=\diag(e^{-u-\eta}\,, e^{u+\eta}) \,, \qquad 
K^{-}(u)=\diag(e^{u}\,, e^{-u})\,, 
\ee 
respectively.\footnote{For the XXX case, we first rescale 
$u \mapsto -i \eta u$ and $R\mapsto \frac{1}{-i\eta}R$ 
before taking the limit $\eta \rightarrow 0$. 
Hence, we have
\be
R(u) = \left(\begin{array}{cccc}
u+i & 0 & 0 & 0\\
0 & u & i & 0 \\
0 & i & u & 0 \\
0 & 0 & 0 & u+i
\end{array} \right) \,,  
\label{RXXX}
\ee
and $K^{+}(u)=K^{-}(u)=\id$.} The transfer matrix commutes for different values of
the spectral parameter
\be
\left[ t(u) \,, t(v) \right] = 0 \,,
\label{commutativity}
\ee
and it contains the Hamiltonian (\ref{Hamiltonian}) 
\footnote{For the case $p=2$ (i.e., $\eta = i \pi/2$), the 
first derivative of the transfer matrix is proportional to 
the identity matrix; hence, the
Hamiltonian is related to the second derivative of the transfer 
matrix, $H=(-1)^{N} \frac{1}{4}t''(0)$.}
\be
H = \alpha\, t'(0) + \beta\, \id\,,
\label{Htrltn}
\ee 
where
\be 
\alpha = \csch(2\eta) \csch^{2(N-1)}\eta \,, \qquad \beta = 
-(N+1)\cosh \eta + \sech \eta \,.
\ee
By taking higher derivatives of the transfer matrix, we obtain the higher conserved charges, 
which commute with each other by virtue of (\ref{commutativity})
\be
H_{n}=\frac{d^{n}}{du^{n}}t(u)\Big\vert_{u=0}\,, \qquad 
\left[H_{n}\,, H_{m} \right] = 0  \,.
\ee

The transfer matrix has $\qg$ symmetry~\cite{Kulish:1991np}
\be
\left[ t(u)\,, S^{z} \right] = 0 \,, \qquad \left[ t(u)\,, S^{\pm} 
\right] = 0\,,
\label{tqgsym}
\ee
where the $\qg$ generators $S^{z}$ and $S^{\pm}$
are given by
\be
S^z &=& \sum_{k=1}^N S^z_k \,, \qquad S^z_k = \frac{1}{2}\sigma^z_k 
\,,
\non \\
S^\pm &=& \sum_{k=1}^N q^{-(S^z_1 + \cdots + S^z_{k-1})}\ S^\pm_{k} \ 
q^{(S^z_{k+1} + \cdots + S^z_N)} \,, \qquad
S^\pm_k = \frac{1}{2}(\sigma^x_k \pm i \sigma^y_k) \,,
\label{Spm-def}
\ee 
and satisfy (\ref{qsl-rel}). The $\qg$ symmetry of the Hamiltonian 
can therefore be understood as a consequence of the 
symmetry of the transfer matrix (\ref{tqgsym}) and the relation 
(\ref{Htrltn}).
The transfer matrix also has the crossing symmetry~\cite{Mezincescu:1991ag}
\be
t(u) = t(-u-\eta) \,.
\ee

The $A$, $B$, $C$, and $D$ operators of the algebraic Bethe ansatz are obtained
from the operator ${\cal U}$ given by
\be
{\cal U}_{a}(u) = T_{a}(u)\, K^{-}_{a}(u)\, \hat T_{a}(u) 
= \left( \begin{array}{cc}
A(u) & B(u) \\
C(u) & D(u) + \frac{\sinh \eta}{\sinh(2u+\eta)} A(u) 
\end{array} \right) \,,
\ee 
in terms of which the transfer matrix (\ref{transfer}) is given by
\be
t(u) = \tr_{a} K^{+}_{a}(u)\, {\cal U}_{a}(u) \,.
\ee
The Bethe states are defined by
\be
|v_{1} \ldots v_{M} \rangle = \prod_{k=1}^{M} B(v_{k})  |0\rangle\,,
\label{Bethestate}
\ee
where $|0\rangle$ is the reference state with all spins up
\be
|0\rangle = {1\choose 0}^{\otimes N} \,,
\label{reference}
\ee
and $v_{1}\,, \ldots \,, v_{M}$ remain to be specified. The Bethe states satisfy the off-shell 
relation\footnote{Details of this computation can be found in e.g. 
\cite{Belliard:2014fsa}.}
\be
t(u) |v_{1} \ldots v_{M} \rangle = \Lambda(u) 
|v_{1} \ldots v_{M} \rangle + 
\sum_{m=1}^{M} \Lambda_{m}|u, v_{1} \ldots \hat v_{m} 
\ldots v_{M} \rangle \,,
\label{offshell}
\ee
where $\Lambda(u)$ is given by the so-called T-Q 
equation\footnote{For the XXX case, the T-Q equation is
\be
\Lambda(u) = 
\frac{2}{(2u+i)}(u+i)^{2N+1}\frac{Q(u-i)}{Q(u)} +
\frac{2}{(2u+i)}u^{2N+1}\frac{Q(u+i)}{Q(u)} \,, \quad 
Q(u) 
=\prod_{k=1}^{M}\left(u-v_{k}\right)\left(u+v_{k}+i\right).
\label{LambdaXXX}
\ee
}
\be
\Lambda(u) = 
\frac{\sinh(2u+2\eta)}{\sinh(2u+\eta)}\sinh^{2N}(u+\eta)\frac{Q(u-\eta)}{Q(u)} +
\frac{\sinh(2u)}{\sinh(2u+\eta)}\sinh^{2N}(u)\frac{Q(u+\eta)}{Q(u)} \,,
\label{Lambda}
\ee
with
\be
Q(u) 
=\prod_{k=1}^{M}\sinh\left(u-v_{k}\right)\sinh\left(u+v_{k}+\eta\right) = Q(-u-\eta)\,.
\ee 
Moreover,
\be
\lefteqn{\Lambda_{m} = f(u,v_{m})
\Bigg[\sinh^{2N}(v_{m}+\eta)
\prod_{\scriptstyle{k \ne m}\atop \scriptstyle{k=1}}^M
\frac{\sinh(v_{m}-v_{k}-\eta) \sinh(v_{m}+v_{k})}
{\sinh(v_{m}-v_{k}) \sinh(v_{m}+v_{k}+\eta)} 
\non} \\
&& -\sinh^{2N}(v_{m})
\prod_{\scriptstyle{k \ne m}\atop \scriptstyle{k=1}}^M
\frac{\sinh(v_{m}-v_{k}+\eta) 
\sinh(v_{m}+v_{k}+2\eta)}
{\sinh(v_{m}-v_{k}) 
\sinh(v_{m}+v_{k}+\eta)}\Bigg] \,,
\label{Lambdam}
\ee
where 
\be
f(u,v) = \frac{\sinh(2u+2\eta)\sinh(2v) \sinh \eta}
{ \sinh(u-v) \sinh(u+v+\eta) \sinh(2v+\eta)} \,.
\label{fuv}
\ee
It follows from (\ref{offshell}) that the Bethe state $|v_{1} \ldots 
v_{M} \rangle$ (\ref{Bethestate}) is an eigenstate 
of the transfer matrix $t(u)$ in (\ref{transfer}) with eigenvalue $\Lambda(u)$ in (\ref{Lambda})
if all the $\Lambda_{m}$ vanish; i.e., according to (\ref{Lambdam}),
if $v_{1}\,, \ldots \,, v_{M}$ satisfy 
\be
\lefteqn{\sinh^{2N}(v_{m}+\eta)
\prod_{\scriptstyle{k \ne m}\atop \scriptstyle{k=1}}^M
\sinh(v_{m}-v_{k}-\eta) \sinh(v_{m}+v_{k})}\non \\
&&=
\sinh^{2N}(v_{m})
\prod_{\scriptstyle{k \ne m}\atop \scriptstyle{k=1}}^M
\sinh(v_{m}-v_{k}+\eta) 
\sinh(v_{m}+v_{k}+2\eta) \,, \quad m = 1, \ldots, M \,.
\label{BAEv}
\ee
These equations coincide with the Bethe equations (\ref{BAE}) upon identifying
\be
v_{m} = \lambda_{m} - \frac{\eta}{2}\,, \quad m = 1, \ldots, M \,.
\ee 
The result (\ref{energy}) for the energy follows from (\ref{Htrltn}) and 
(\ref{Lambda}).

In passing to (\ref{BAEv}), it was assumed that the factor 
$f(u,v_{m})$ in (\ref{Lambdam}) is regular. However, $f(u,v)$ has a pole at $v=-\eta/2$,
as can be seen from (\ref{fuv}).
Hence, solutions of the Bethe equations (\ref{BAEv}) containing  
$v_{m}=-\eta/2$ must be discarded, since $\Lambda_{m}$ will not 
vanish, and therefore the corresponding Bethe state will not be an 
eigenstate of the transfer matrix. Similarly, $v_{m}=-\eta/2 + i 
\pi/2$ must be excluded.

In other words, solutions of the Bethe equations (\ref{BAE}) with
$\lambda_{m}=0$ or $\lambda_{m}=i \pi/2$ must be discarded, because
they do not correspond to eigenstates of the transfer matrix. It has 
also been argued \cite{Fendley:1994cz} that such solutions should be
discarded because the corresponding coordinate Bethe ansatz wave function
\cite{Alcaraz:1987uk} vanishes identically.

For generic values of $q$, the on-shell (i.e., with Bethe 
equations satisfied) Bethe state (\ref{Bethestate}) 
is an $\qg$ highest-weight state \cite{Pasquier:1989kd, Hou:1991tc, Mezincescu:1991rb}
\be
S^{+} |v_{1} \ldots v_{M} \rangle = 0 \,,
\label{highestweight}
\ee
with
\be
S^{z} |v_{1} \ldots v_{M} \rangle = \left(\frac{N}{2}-M\right)  |v_{1} \ldots 
v_{M} \rangle \,.
\label{Szeig}
\ee 
The on-shell Bethe state (\ref{Bethestate}) is therefore an 
eigenstate of the Casimir operator (see e.g.~\cite{Pasquier:1989kd})
\be
S^{2} = S^{-} S^{+} + \left(\left[S^{z}+\tfrac{1}{2}\right]_{q} \right)^{2} 
- \left[\tfrac{1}{2}\right]_{q}^{2} \,,
\label{Casimir}
\ee
with corresponding eigenvalue
\be
S^{2} |v_{1} \ldots v_{M} \rangle = \left(\left[j+\tfrac{1}{2}\right]_{q}^{2} - 
 \left[\tfrac{1}{2}\right]_{q}^{2} \right)|v_{1} \ldots v_{M} \rangle \,,
 \ee 
where $[x]_q$ is defined in~\eqref{qsl-rel} and  the spin $j$ is given by
\be
j=\frac{N}{2}-M \,.
\label{spinj}
\ee
The requirement $j \ge 0$ implies that $M \le \frac{N}{2}$. 
The lower-weight states ($S^{z} < j$) of the spin-$j$  representation of $\qg$
can be obtained by repeatedly acting on the highest-weight state ($S^{z} = j$) by the $S^{-}$ operator defined in~\eqref{Spm-def}.

\section{Temperley-Lieb algebra}\label{sec:TL}

The Hamiltonian (\ref{Hamiltonian}) can evidently be re-expressed (up to an 
additive constant) as~\cite{Pasquier:1989kd} 
\be
H = -2\sum_{k=1}^{N-1} e_{k} \,,
\ee
where the $e_{k}$ are given by
\be
e_{k} = - {1\over 2}\left(\sigma^x_k \sigma^x_{k+1} +
\sigma^y_k \sigma^y_{k+1} \right) - {1\over 4}( q + q^{-1})\left(
\sigma^z_k \sigma^z_{k+1} - 1 \right)
+ {1\over 4}( q - q^{-1})
\left( \sigma^z_k - \sigma^z_{k+1} \right) \,.
\ee
The $e_{k}$ can be shown to satisfy the Temperley-Lieb algebra~\cite{Temperley:1971iq}
\be
e_{k}^{2} &=& \delta e_{k}\,, \non \\
e_{k} e_{k \pm 1} e_{k}  &=& e_{k} \,, \non \\
e_{k} e_{j} &=& e_{j} e_{k} \,, \quad |j-k|>1 \,,
\label{TL}
\ee
where $\delta$ (the so-called fugacity or loop parameter) is given by
\be
\delta = q + q^{-1}  \,.
\ee
For $q=e^{i \pi/p}$, it follows that $\delta=2\cos{\frac{\pi}{p}}$. 
For all values of $q$, including 
the roots of unity, the Temperley-Lieb algebra is identified with the 
maximum algebra commuting with (or centralizer of) $\qg$, 
see~\cite{Martin:1991pk,Martin:1991zk}. 

\section{Bethe solutions at $p=2$}
\label{sec:p2}

The case $p=2$ (i.e., $\eta = i \pi/2$) is sufficiently simple to be 
analyzed analytically. The Bethe equations (\ref{BAEx}) decouple and reduce to
\be
\left(\frac{i x_{k} -1}{x_{k} - i}\right)^{2N} = 1 \,,
\ee
since $q^{2}=-1$ and therefore the terms with $\prod_{j \ne k}$ 
cancel. It follows that
\be
\frac{i x_{k} -1}{x_{k} - i} = e^{i \omega_{l}}\,, \qquad \omega_{l} = \frac{2\pi l}{2N}\,, \qquad l = 0, 1, \ldots, 2N-1,
\ee
and therefore
\be
\lambda_{k} = \frac{1}{2} \ln x_{k} = \frac{1}{2} \ln \left( 
\frac{i e^{i \omega_{l}} -1 }{e^{i \omega_{l}}  - i} \right) \,.
\label{lambdap2}
\ee 
The admissible Bethe roots (recall (\ref{admissibleXXZa}) and 
(\ref{admissibleXXZb})) appear as
$(\lambda\,, \lambda + i \pi/2)$ (i.e.,  pairs of roots that differ by 
$i\pi/2$) corresponding to the following pairs of $l$ values
\be
(l\,, N - l)\,, \qquad l = 1, 2, \ldots , l_{\rm max} \,,
\ee
where 
\be
l_{\rm max} = \left\{ \begin{array}{cc}
                        (N-2)/2 & \quad N = {\rm even} \\
			(N-1)/2 & \quad N = {\rm odd} 
			\end{array} \right. \,.
\ee
It follows that the number of solutions ${\cal N}(N,M)$ for $M=1$ is given by
\be
{\cal N}(N,1) = 2 l_{\rm max} =  \left\{ \begin{array}{cc}
                       N-2 & \quad N = {\rm even} \\
		       N-1 & \quad N = {\rm odd} 
			\end{array} \right. \,.
\ee

In order to construct Bethe states (\ref{Bethestate}) with $M>1$, one would naively expect to 
be able to choose any $M$ roots from the ${\cal N}(N,1)$ admissible roots.
However, a Bethe vector with two roots that differ 
by $i\pi/2$ is an exact complete 2-string (\ref{exactcompleterstring}).
Hence, any solution of the Bethe equations that 
contains a pair of roots that differ by $i\pi/2$ is not admissible.  Since the 
${\cal N}(N,1)$ admissible roots all come in pairs that differ by 
$i\pi/2$, it follows that the number of solutions for $M\ge 1$ is given by (notice 
the double factorials)
\be
{\cal N}(N,M) = \frac{{\cal N}(N,1)!!}{M! ({\cal N}(N,1)-2M)!!} = 
 \left\{ \begin{array}{cc}
                       \frac{(N-2)!!}{M! (N-2-2M)!!} & \quad N = {\rm even} \\
		       \frac{(N-1)!!}{M! (N-1-2M)!!} & \quad N = {\rm odd} 
			\end{array} \right. \,.
			\label{p2results}
\ee 
The results for $N=2, \ldots, 9$ 
are displayed in Table \ref{table:XXZp} (a).

\section{Explanations of deviations}\label{sec:deviations}

We consider here in detail the cases of the conjecture 
(\ref{mainconjecture2})
for which ${\cal N}(N,M) < d^{0}_{\frac{N}{2}-M}$,
and the cases of the conjecture (\ref{dimconjecture}) for which
${\cal D}(N,M) > \dim T_{\frac{N}{2}-M}$.  
We argue that these deviations occur when two
or more tilting modules become degenerate. 
The idea is that we count the total degeneracies of
generalized eigenvalues of the transfer matrix; and by comparing them with dimensions of the
tilting modules  
and using the $S^z$ values for corresponding generalized eigenstates, we infer which tilting modules are
degenerate. As we do not construct a basis in these tilting
modules explicitly, our arguments are rather indirect 
but definitive.

We analyze below only the cases $p=2$ and $p=3$, because for higher values of $p$ we would need to go beyond $N=10$ 
which 
exceeds the capabilities of our 
available computer resources.
For $p=2$, we use known facts on representation theory
of both the Temperley--Lieb and $U_q sl(2)$ 
algebras~\cite{Martin:1991pk,Read:2007qq,Gainutdinov:2012qy}:

\begin{enumerate}
\item For odd $N$, the TL algebra is semisimple and the Hamiltonian
is diagonalizable -- it is the only semisimple/diagonalizable case at roots
of unity. Hence, for this case all the eigenvalues and eigenvectors  
are ordinary (i.e. not generalized). For even $N$, the TL algebra is non-semisimple and the Hamiltonian has Jordan blocks of maximum rank~$2$.

\item For odd $N$, the tilting $U_q sl(2)$-modules $T_j$
in~\eqref{decomposition} appear for half-integer $j$ and are
irreducible.  The $S^z$ spectrum is then usual one $\{j, j-1,\ldots,
-j\}$.  For even $N$, each tilting $U_q sl(2)$-module $T_j$, where $j$
is a positive integer, is indecomposable but reducible and is
composed of the spin-$j$ and the spin-$(j-1)$ modules (recall the
discussion above~\eqref{dimTj}), where each spin-$j$ module is also
reducible but indecomposable and has the unique submodule isomorphic
to the head (or irreducible quotient) of the spin-$(j-1)$ module.  The
dimension of the head of the spin-$j$ module is $j+1$ and we denote
the head by $\Irr{j}$.  In total, the sub-quotient structure of $T_j$
in terms of the irreducible modules $\Irr{j}$ is 
\begin{equation}
\xymatrix@R=22pt@C=1pt{
\mbox{}&\\
&T_j\quad:\
\mbox{}&\\
}
\xymatrix@R=22pt@C=10pt@W=4pt@M=6pt{
&\Irr{j-1}\ar[dl]\ar[dr]&\\
\Irr{j-2}\ar[dr]
&&\mbox{}\;\Irr{j}\;\;\;\;\;\ar[dl]\\
&\Irr{j-1}&
}
\label{Tj-diag}
\end{equation}
where arrows correspond to
irreversible action of $U_q sl(2)$ generators and we set $\Irr{-1}=0$.
In the decomposition~\eqref{decomposition}, a direct summand $T_j$ has $S^z$ spectrum $\{j, 2\times (j-1),  2\times (j-2), \ldots,  2\times (-j+1), -j\}$ while each irreducible sub-quotient $\Irr{j}$ has $S^z$ spectrum $\{j,j-2,\ldots,-j+2,-j\}$.
We also note that it is only the states in the head of $T_j$ -- the 
top sub-quotient $\Irr{j-1}$ in~\eqref{Tj-diag} -- on which the Hamiltonian is non-diagonalizable.
\end{enumerate}

\subsection{$p =2\,, \ N= 5$}

For $p=2$ and  $N=5$, the decomposition (\ref{decomposition}) into tilting modules 
 is given by 
\be
5 T_{\frac{1}{2}} \oplus 4 T_{\frac{3}{2}} \oplus T_{\frac{5}{2}} \,. \non 
\ee
We claim that $T_{\frac{5}{2}}$ and one of the $T_{\frac{1}{2}}$ are 
degenerate, and therefore $n_{\frac{5}{2},\frac{1}{2}}=1$, 
$n_{\frac{1}{2}}=1$  (all others are zero). 
Indeed, the subspace with energy $E=0$ \footnote{Strictly
speaking, we should consider (generalized) eigenvalues of the transfer matrix.  We
consider in this appendix instead (generalized) eigenvalues of the Hamiltonian, which are easier
to report and give the same results.}, which includes the reference state ($M=0\,, 
j=\frac{5}{2}$) belonging to $T_{\frac{5}{2}}$, can be shown to have dimension~$8$.
That is, the number of $E=0$ eigenstates is $8$,
which is the sum 
of dimensions of $T_{\frac{5}{2}}$ (6) and $T_{\frac{1}{2}}$ (2), 
recall~\eqref{dimTj}. 
This implies that
\be
{\cal D}(5,0) &=& \dim T_{\frac{5}{2}} +  n_{\frac{5}{2},\frac{1}{2}} 
\dim T_{\frac{1}{2}} = 6 + 1*2 =8 
\,,
\non \\
{\cal N}(5,2) &=& d_{\frac{1}{2}}^{0} - n_{\frac{1}{2}} = 5 - 1 = 4 
\,,
\ee
in agreement with Tables~\ref{table:XXZpmore} and~\ref{table:XXZp}, respectively.

\subsection{$p =2\,, \ N= 6$}

For $p=2, N=6$, the decomposition (\ref{decomposition}) into tilting modules is given by
\be
5 T_1 \oplus 4 T_2 \oplus T_3 \,. \non 
\ee
We claim that $T_3$ and one of the  $T_1$ are 
degenerate, and therefore $n_{3,1}=1$, $n_{1}=1$  (all others are zero).
Indeed, the subspace with 
generalized $H$-eigenvalue $E=0$, which includes the reference state ($M=0\,, 
j=3$) belonging to $T_3$, can be shown to have dimension $16$, which is the sum 
of dimensions of $T_3$ (12) and $T_1$ (4). 
This implies that 
\be
{\cal D}(6,0) &=& \dim T_{3} +  n_{3,1} 
\dim T_{1} = 12 + 1*4 = 16\,, \non \\
{\cal N}(6,2) &=& d_{1}^{0} - n_{1} = 5 - 1 = 4 
\,.
\ee

\subsection{$p =2\,, \ N= 7$}

For $p=2, N=7$, the decomposition (\ref{decomposition}) into tilting modules is given by
\be
14 T_{\frac{1}{2}} \oplus 14 T_{\frac{3}{2}} \oplus 6 T_{\frac{5}{2}} \oplus  
T_{\frac{7}{2}} \,. \non 
\ee
We claim that $T_{\frac{7}{2}}$ and two of the  $T_{\frac{3}{2}}$ are 
degenerate, and therefore $n_{\frac{7}{2},\frac{3}{2}}=2$, 
$n_{\frac{3}{2}}=2$. 
Indeed, the subspace with energy $E=0$, which includes the reference state ($M=0\,, 
j=\frac{7}{2}$), can be shown to have dimension 16,  which  
can be now  either the sum 
of dimensions of $T_{\frac{7}{2}}$ (8) and two $T_{\frac{3}{2}}$ (2*4=8) or as $ \dim
T_{\frac{7}{2}}+ \dim T_{\frac{5}{2}} + \dim T_{\frac{1}{2}}$ or the sum $\dim T_{\frac{5}{2}} + \dim T_{\frac{3}{2}} +
2\dim T_{\frac{1}{2}}$.  Looking at $S^z$-sectors for
these $16$ eigenstates:
\be
S^z=\Bigl\{
\pm\ffrac{7}{2}, \pm\ffrac{5}{2}, 3\times \bigl(\pm\ffrac{3}{2}\bigr),
3\times \bigl(\pm\ffrac{1}{2}\bigr)
\Bigr\} \non
\ee
and recalling the discussion above~\eqref{Tj-diag}, we  identify
precisely the tilting modules they belong to as $T_{\frac{7}{2}}\oplus 2T_{\frac{3}{2}}$.
This gives 
\be
{\cal D}(7,0) &=& \dim T_{\frac{7}{2}} +  n_{\frac{7}{2},\frac{3}{2}} 
\dim T_{\frac{3}{2}} = 8 + 2*4 = 16 \,, \non \\
{\cal N}(7,2) &=& d_{\frac{3}{2}}^{0} - n_{\frac{3}{2}} = 14 - 2 = 12 
\,.
\ee
Moreover, we claim that each of the $6$ $T_{\frac{5}{2}}$ are 
degenerate with $6$ $T_{\frac{1}{2}}$, and therefore 
$n_{\frac{5}{2},\frac{1}{2}}=1$, 
$n_{\frac{1}{2}}=6$. 
Indeed, we find that there are $6$ 
energy eigenvalues 
(namely, $\pm 3.60388$, $\pm 2.49396$, and $\pm 0.890084$)
that are each $8$-fold degenerate; while 
${\rm dim}\ T_{\frac{5}{2}} = 6$ and ${\rm dim}\ T_{\frac{1}{2}} = 
2$. This implies that 
\be
{\cal D}(7,1) &=& \dim T_{\frac{5}{2}} +  n_{\frac{5}{2},\frac{1}{2}} 
\dim T_{\frac{1}{2}} = 6 + 1*2 = 8 \,, \non \\
{\cal N}(7,3) &=& d_{\frac{1}{2}}^{0} - n_{\frac{1}{2}} = 14 - 6 = 8 \,.
\ee

\subsection{$p =2\,, \ N= 8$}

For $p=2, N=8$, the decomposition (\ref{decomposition}) into tilting modules is given by
\be
14 T_1 \oplus 14 T_2 \oplus 6 T_3 \oplus T_4 \,. \non 
\ee
We claim that $T_4$ and two of the  $T_2$ are 
degenerate, and therefore $n_{4,2}=2$, $n_{2}=2$. 
Indeed, the subspace with 
generalized $H$-eigenvalue $E=0$, which includes the reference state ($M=0\,, 
j=4$), can be shown to have dimension 32, which is either the sum 
of dimensions of $T_4$ (16) and two $T_2$ (2*8=16)
or one of these sums $ \dim T_{4} + \dim T_{3}+ \dim T_{1}= \dim T_{4} + 4\dim T_{1}= \dim T_{4} + \dim T_{2}+2 \dim T_{1}$. 
 Looking then at $S^z$-sectors for
these $32$ generalized eigenstates,
we find that the 24 $S^z$ eigenvalues corresponding to 
ordinary eigenvectors are
\be
S^z=\Bigl\{
\pm4, \pm3, 4\times (\pm2),
3\times (\pm1), 6\times 0
\Bigr\} \non
\ee
and the 8 $S^z$ eigenvalues corresponding to 
generalized eigenvectors are
\be
S^z=\Bigl\{
\pm3, 
3\times (\pm1)
\Bigr\}. \non
\ee
Recalling the discussion about tilting modules and their $S^z$ spectrum (above~\eqref{Tj-diag}) we  identify
precisely the tilting modules the $32$ generalized eigenstates belong to as $T_{4}\oplus 2T_{2}$.
So, the $S^z$ spectrum we found implies 
\be
{\cal D}(8,0) &=& \dim T_{4} +  n_{4,2} 
\dim T_{2} = 16 + 2*8 = 32\,, \non \\
{\cal N}(8,2) &=& d_{2}^{0} - n_{2} = 14 - 2 = 12 \,.
\ee
Moreover, we claim that each of the 6 $T_3$ are 
degenerate with 6 $T_1$, and therefore $n_{3,1}=1$, $n_{1}=6$. 
Indeed, we find that there are 6 
energy eigenvalues 
(namely, $\pm 2\sqrt{2+\sqrt{2}}$, $\pm 2\sqrt{2}$, $\pm  2\sqrt{2-\sqrt{2}}$)
that are each 16-fold degenerate; while 
${\rm dim}\ T_3 = 12$ and ${\rm dim}\ T_1 = 
4$. This implies that 
\be
{\cal D}(8,1) &=& \dim T_{3} +  n_{3,1} 
\dim T_{1} = 12 + 1*4 = 16\,, \non \\
{\cal N}(8,3) &=& d_{1}^{0} - n_{1} = 14 - 6 = 8 \,.
\ee

\subsection{$p =2\,, \ N= 9$} 
For $p=2$ and $N=9$, the decomposition (\ref{decomposition}) into tilting modules is given by
\be
42 T_{\frac{1}{2}} \oplus 48 T_{\frac{3}{2}} \oplus 27 T_{\frac{5}{2}} \oplus  
8 T_{\frac{7}{2}} \oplus  
T_{\frac{9}{2}} \,. \non 
\ee
Using analysis similar to the previous cases, we claim that the nonzero $n_{jk}$ are
\be
n_{\frac{9}{2},\frac{5}{2}}=3\,, \quad 
n_{\frac{9}{2},\frac{1}{2}}=2\,, \quad
n_{\frac{7}{2},\frac{3}{2}}=2\,, \quad
n_{\frac{5}{2},\frac{1}{2}}=1\,,
\label{njkp2n9}
\ee
and the nonzero $n_{j}$ are
\be
n_{\frac{5}{2}}=3\,, \quad 
n_{\frac{3}{2}}=16\,, \quad 
n_{\frac{1}{2}}=26\,.
\label{njp2n9}
\ee
Hence,
\be
{\cal D}(9,0) &=& \dim T_{\frac{9}{2}} +  
n_{\frac{9}{2},\frac{5}{2}}  \dim T_{\frac{5}{2}} +  
n_{\frac{9}{2},\frac{1}{2}}  \dim T_{\frac{1}{2}} = 10 + 3*6 + 2*2 = 
32\,, \non \\
{\cal D}(9,1) &=& \dim T_{\frac{7}{2}} +  
n_{\frac{7}{2},\frac{3}{2}}  \dim T_{\frac{3}{2}} = 8 +  2*4 = 
16\,, \non \\
{\cal D}(9,2) &=& \dim T_{\frac{5}{2}} +  
n_{\frac{5}{2},\frac{1}{2}}  \dim T_{\frac{1}{2}} = 6 +  1*2 = 
8\,, 
\ee
and
\be
{\cal N}(9,2) &=& d_{\frac{5}{2}}^{0} - n_{\frac{5}{2}} = 27 - 3 = 24 \,, \non \\
{\cal N}(9,3) &=& d_{\frac{3}{2}}^{0} - n_{\frac{3}{2}} = 48 - 16 = 32 \,, \non \\
{\cal N}(9,4) &=& d_{\frac{1}{2}}^{0} - n_{\frac{1}{2}} = 42 - 26 = 16 \,,
\ee
in agreement with Tables \ref{table:XXZpmore} and \ref{table:XXZp}, 
respectively. 
One can also verify that the $n_{j}$'s (\ref{njp2n9}) can be obtained 
from the $n_{jk}$'s (\ref{njkp2n9}) using (\ref{njnjk}).

\subsection{$p =3\,, \ N= 8$}

For $p=3, N=8$, the decomposition (\ref{decomposition}) into tilting modules is given by
\be
T_0 \oplus 28 T_1 \oplus 13 T_2\oplus 7 T_3 \oplus T_4 \,. \non 
\ee
We claim that $T_4$ and one of the  $T_1$ are 
degenerate, and therefore $n_{4,1}=1$, $n_{1}=1$  (all others are zero). 
Indeed, the subspace with energy $E=0$, which includes the reference state ($M=0\,, 
j=4$), can be shown to have dimension 12, which is the sum 
of dimensions of $T_4$ (9) and $T_1$ (3). 
This implies that
\be
{\cal D}(8,0) &=& \dim T_{4} +  n_{4,1} 
\dim T_{1} = 9 + 1*3 = 12\,, \non \\
{\cal N}(8,3) &=& d_{1}^{0} - n_{1} = 28 - 1 = 27 \,.
\ee

\section{Numerical results}\label{sec:numerical}

Our numerical solutions of the Bethe equations up to  $N=8$ are presented in
Tables \ref{table:XXXsltns}-\ref{table:XXZp5sltns2}.  
These results were obtained using homotopy continuation \cite{BHSW1} (see also
\cite{Hao:2013jqa} and references therein for further details). 

\newpage

\begin{table}
\tiny
\centering
\begin{tabular}{|c|c|c| c c c|}\hline
N&M&number  &$\lambda_{1}$&$\lambda_{2}$&$\lambda_{3}$\\\hline
\multirow{1}{*}{2}&\multirow{1}{*}{1}&1&0.5&&\\\hline
\multirow{2}{*}{3}&\multirow{2}{*}{1}&1&0.8660254037844386&&\\\cline{3-6}&&2&0.2886751345948129&&\\\hline
\multirow{3}{*}{4}&\multirow{3}{*}{1}&1&1.207106781186547&&\\\cline{3-6}&&2&0.5&&\\\cline{3-6}&&3&0.2071067811865475&&\\\hline
\multirow{3}{*}{4}&\multirow{3}{*}{2}&1&0.7160149594491338&0.7160149594491338&\\&&&+0.5125206553446844$i$&-0.5125206553446844$i$&\\\cline{3-6}&&2&0.6683262276726571&0.2309546565991595&\\\hline
\multirow{4}{*}{5}&\multirow{4}{*}{1}&1&1.538841768587627&&\\\cline{3-6}&&2&0.6881909602355868&&\\\cline{3-6}&&3&0.3632712640026804&&\\\cline{3-6}&&4&0.1624598481164532&&\\\hline
\multirow{7}{*}{5}&\multirow{7}{*}{2}&1&1.115042120183109&1.115042120183109&\\&&&+0.5450541101265968$i$&-0.5450541101265968$i$&\\\cline{3-6}&&2&0.9704069411911774&0.1723800721632705&\\\cline{3-6}&&3&0.5137119304322965&0.5137119304322965&\\&&&+0.4996020494993916$i$&-0.4996020494993916$i$&\\\cline{3-6}&&4&0.9496686956332134&0.3969680639294287&\\\cline{3-6}&&5&0.4272945057154192&0.1793374003754359&\\\hline
\multirow{5}{*}{6}&\multirow{5}{*}{1}&1&1.866025403784439&&\\\cline{3-6}&&2&0.2886751345948129&&\\\cline{3-6}&&3&0.1339745962155613&&\\\cline{3-6}&&4&0.5&&\\\cline{3-6}&&5&0.8660254037844386&&\\\hline
\multirow{12}{*}{6}&\multirow{12}{*}{2}&1&1.234440793585582&0.5389490693006668&\\\cline{3-6}&&2&0.8418003199559988&0.8418003199559988&\\&&&+0.4947462450429116$i$&-0.4947462450429116$i$&\\\cline{3-6}&&3&0.3905082158626772&0.3905082158626772&\\&&&+0.5000053666355159$i$&-0.5000053666355159$i$&\\\cline{3-6}&&4&1.277389814218266&0.1387104764546264&\\\cline{3-6}&&5&1.471796355306884&1.471796355306884&\\&&&-0.5824072783212229$i$&+0.5824072783212229$i$&\\\cline{3-6}&&6&0.591788951573015&0.3152209587092826&\\\cline{3-6}&&7&0.1457831570066063&0.3239416643695967&\\\cline{3-6}&&8&0.5975749352330829&0.1434644688717632&\\\cline{3-6}&&9&1.266274529052914&0.3019322716047725&\\\hline\multirow{10}{*}{6}&\multirow{10}{*}{3}&1&0.7487200726653173&0.5881061192792989&0.5881061192792989\\&&&&+0.5011583393895944$i$&-0.5011583393895944$i$\\\cline{3-6}&&2&0.9677400136112142&0.9476918141366062&0.9476918141366062\\&&&&+0.9956807427853811$i$&-0.9956807427853811$i$\\\cline{3-6}&&3&0.774814166699722&0.35438
 89174298362&0.1551499348511761\\\cline{3-6}&&4&0.7901794336920558&0.148626658019744&0.7901794336920558\\&&&+0.5103219367879035$i$&&-0.5103219367879035$i$\\\cline{3-6}&&5&0.7601147488943615&0.3341849467072039&0.7601147488943615\\&&&+0.5085412675384237$i$&&-0.5085412675384237$i$\\\hline\end{tabular}
\caption{\small The admissible solutions of the XXX Bethe
   equations (\ref{BAEXXX}) up to $N=6$.}
\label{table:XXXsltns}
\end{table}

\begin{table}
\tiny \centering
\begin{tabular}{|c|c|c| c c c|}\hline
N&M&number  &$x_{1}$&$x_{2}$&$x_{3}$\\\hline
\multirow{1}{*}{2}&\multirow{1}{*}{1}&1&0.9950207489532265&&\\&&&+0.0996679946249558$i$&&\\\hline
\multirow{4}{*}{3}&\multirow{4}{*}{1}&1&0.9851362571667408&&\\&&&+0.1717747211189849$i$&&\\\cline{3-6}&&2&0.9983374903000208&&\\&&&+0.05763900989309033$i$&&\\\hline\multirow{6}{*}{4}&\multirow{6}{*}{1}&1&0.9713235174298164&&\\&&&+0.2377616968474302$i$&&\\\cline{3-6}&&2&0.9950207489532265&&\\&&&+0.09966799462495615$i$&&\\\cline{3-6}&&3&0.9991439299596814&&\\&&&+0.04136915789236264$i$&&\\\hline\multirow{4}{*}{4}&\multirow{4}{*}{2}&1&0.9911071329262735&0.9989343649801794&\\&&&+0.1330663408329167$i$&+0.04615337974239554$i$&\\\cline{3-6}&&2&1.096607956599227&1.096607956599227&\\&&&+0.1576690437914444$i$&-0.1576690437914444$i$&\\\hline\multirow{8}{*}{5}&\multirow{8}{*}{1}&1&0.9538101147863981&&\\&&&+0.3004101611649615$i$&&\\\cline{3-6}&&2&0.9905881296732647&&\\&&&+0.1368764309529703$i$&&\\\cline{3-6}&&3&0.9973685392726434&&\\&&&+0.07249825424900729$i$&&\\\cline{3-6}&&4&0.9994731533039727&&\\&&&+0.03245636801328048$i$&&\\\hline\multirow{10}{*}{5}&\multirow{10}{*}{2}&1&1.087533505007
 09&1.08753350500709&\\&&&+0.2455058557178879$i$&-0.2455058557178879$i$&\\\cline{3-6}&&2&0.9821261824941047&0.996853923761068&\\&&&+0.1882237011100266$i$&+0.07926067550912498$i$&\\\cline{3-6}&&3&1.099298174842604&1.099298174842604&\\&&&+0.1129555095638716$i$&-0.1129555095638716$i$&\\\cline{3-6}&&4&0.9813430842510684&0.9994062436043682&\\&&&+0.1922647939499075$i$&+0.03445519183818567$i$&\\\cline{3-6}&&5&0.9963579374226493&0.9993575911189618&\\&&&+0.08526934111909155$i$&+0.03583859752984044$i$&\\\hline\multirow{10}{*}{6}&\multirow{10}{*}{1}&1&0.9328105645067954&&\\&&&+0.3603671055250655$i$&&\\\cline{3-6}&&2&0.9851362571667408&&\\&&&+0.1717747211189849$i$&&\\\cline{3-6}&&3&0.9983374903000208&&\\&&&+0.05763900989309033$i$&&\\\cline{3-6}&&4&0.9950207489532265&&\\&&&+0.09966799462495615$i$&&\\\cline{3-6}&&5&0.9996416778201899&&\\&&&+0.02676781583984422$i$&&\\\hline\multirow{18}{*}{6}&\multirow{18}{*}{2}&1&1.075224471519206&1.075224471519206&\\&&&+0.3236217054614372$i$&-0.3236217054
 614372$i$&\\\cline{3-6}&&2&0.9699960698659229&0.9942054556525597&\\&&&+0.2431205965044173$i$&+0.1074965671576824$i$&\\\cline{3-6}&&3&0.9930254383362942&0.9980164632164529&\\&&&+0.1179002918444679$i$&+0.06295346812466036$i$&\\\cline{3-6}&&4&1.10182556643507&1.10182556643507&\\&&&+0.08593996563356507$i$&-0.08593996563356507$i$&\\\cline{3-6}&&5&0.9684452798485436&0.9981790461343699&\\&&&+0.2492262826009249$i$&+0.06032074152627307$i$&\\\cline{3-6}&&6&1.088602293651335&1.088602293651335&\\&&&+0.1841309747335413$i$&-0.1841309747335413$i$&\\\cline{3-6}&&7&0.9979056525281584&0.9995755201710089&\\&&&+0.06468623232458623$i$&+0.02913382012124421$i$&\\\cline{3-6}&&8&0.9678938580855082&0.9996155034837294&\\&&&+0.2513592637647361$i$&+0.02772805790115951$i$&\\\cline{3-6}&&9&0.9928890635454395&0.9995888688063166&\\&&&+0.1190433009113084$i$&+0.02867217045339162$i$&\\\hline\multirow{10}{*}{6}&\multirow{10}{*}{3}&1&1.094320030741045&0.9977671423426131&1.094320030741045\\&&&+0.167153344511961$i
 $&+0.06678869411401678$i$&-0.167153344511961$i$\\\cline{3-6}&&2&1.093678036321473&0.9995582494860389&1.093678036321473\\&&&+0.1737932036777591$i$&+0.02972046238546064$i$&-0.1737932036777591$i$\\\cline{3-6}&&3&0.9815106975959883&1.19864337246897&1.19864337246897\\&&&+0.1914072895807168$i$&+0.2289583093421947$i$&-0.2289583093421947$i$\\\cline{3-6}&&4&0.9888241361234739&1.097821757712908&1.097821757712908\\&&&+0.1490866453431209$i$&+0.1293812108099802$i$&-0.1293812108099802$i$\\\cline{3-6}&&5&0.9880520841223098&0.9974912917057552&0.9995188710316143\\&&&+0.1541203395453045$i$&+0.07078928570895365$i$&+0.03101655125392326$i$\\\hline\end{tabular}
\caption{\small The admissible solutions of the XXZ Bethe equations
   (\ref{BAEx}) with $\eta=0.1$ up to $N=6$.} \label{table:XXZsltns}
\end{table}

\begin{table}
\tiny \centering
\begin{tabular}{|c|c|c| c c c|}\hline
N&M&number  &$x_{1}$&$x_{2}$&$x_{3}$\\\hline
\multirow{1}{*}{2}&\multirow{1}{*}{1}&1&3.732050807568877&&\\\hline
\multirow{1}{*}{3}&\multirow{1}{*}{1}&1&2&&\\\hline
\multirow{3}{*}{4}&\multirow{3}{*}{1}&1&1.628626279736931&&\\\cline{3-6}&&2&-6.078116022520112&&\\\cline{3-6}&&3&3.732050807568876&&\\\hline
\multirow{1}{*}{4}&\multirow{1}{*}{2}&1&5.551933372263207&1.668669261292973&\\\hline
\multirow{4}{*}{5}&\multirow{4}{*}{1}&1&1.461818651603003&&\\\cline{3-6}&&2&2.445124904035096&&\\\cline{3-6}&&3&-3.574329190217507&&\\\cline{3-6}&&4&8.73968131822042&&\\\hline
\multirow{1}{*}{5}&\multirow{1}{*}{2}&1&1.495162537164605&2.744357057158133&\\\hline
\multirow{4}{*}{6}&\multirow{4}{*}{1}&1&1.366025403784439&&\\\cline{3-6}&&2&2&&\\\cline{3-6}&&3&-2.732050807568878&&\\\cline{3-6}&&4&3.732050807568876&&\\\hline
\multirow{10}{*}{6}&\multirow{10}{*}{2}&1&2.121740590131168&1.389071336983916&\\\cline{3-6}&&2&2.065764252710569&4.618645451098817&\\\cline{3-6}&&3&4.708465577106654&1.37980190200904&\\\cline{3-6}&&4&1.94185325885711&-6.956024177556002&\\\cline{3-6}&&5&7.06963925607305&-3.289527548618499&\\\cline{3-6}&&6&2.033456364524103&2.033456364524103&\\&&&-3.572707112158111$i$&+3.572707112158111$i$&\\\cline{3-6}&&7&-5.515508983877778&3.211606419543336&\\\cline{3-6}&&8&-11.72375959717719&-2.585294089372629&\\\cline{3-6}&&9&-7.397095726637462&1.354263271640707&\\\hline
\multirow{1}{*}{6}&\multirow{1}{*}{3}&1&2.184002431728489&7.103070621468425&1.399963454482446\\\hline
\multirow{6}{*}{7}&\multirow{6}{*}{1}&1&1.303554144675824&&\\\cline{3-6}&&2&1.770225971730896&&\\\cline{3-6}&&3&-2.307585402462597&&\\\cline{3-6}&&4&2.706596741879841&&\\\cline{3-6}&&5&-11.05630589583649&&\\\cline{3-6}&&6&6.245703131914746&&\\\hline\multirow{13}{*}{7}&\multirow{13}{*}{2}&1&1.836863203195035&1.319993390462937&\\\cline{3-6}&&2&2.970622089446266&1.818186625013599&\\\cline{3-6}&&3&1.316313662306279&2.990958243918946&\\\cline{3-6}&&4&9.876567762581274&2.793117991352707&\\\cline{3-6}&&5&1.288545404441192&-4.146370060660599&\\\cline{3-6}&&6&-3.574329190217507&2.445124904035096&\\\cline{3-6}&&7&4.053394950868378&-2.890714726422575&\\\cline{3-6}&&8&-2.178060016150124&-6.266169856419983&\\\cline{3-6}&&9&1.547869212009354&1.547869212009354&\\&&&+2.679110191382225$i$&-2.679110191382225$i$&\\\cline{3-6}&&10&-3.978767068330315&1.712236436628736&\\\cline{3-6}&&11&10.50224778955699&1.308224352206094&\\\cline{3-6}&&12&1.788230153989182&10.3328199011193&\\\cline{3-6}&&13&-2.40
 322747023718&14.47572976817471&\\\hline\multirow{1}{*}{7}&\multirow{1}{*}{3}&1&1.332957726595594&3.390479290658175&1.890753798751596\\\hline\end{tabular}
\caption{\small The admissible solutions of the XXZ Bethe equations
   (\ref{BAEx}) with $q=e^{i\pi/3}$ up to $N=7$.} \label{table:XXZp3sltns1}
\end{table}

\begin{table}
\tiny \centering
\begin{tabular}{|c|c|c| c c c|}\hline
N&M&number  &$x_{1}$&$x_{2}$&$x_{3}$\\\hline
\multirow{7}{*}{8}&\multirow{7}{*}{1}&1&1.259483895472751&&\\\cline{3-6}&&2&-2.051228571072364&&\\\cline{3-6}&&3&2.256125795901561&&\\\cline{3-6}&&4&3.732050807568876&&\\\cline{3-6}&&5&1.628626279736931&&\\\cline{3-6}&&6&-6.078116022520112&&\\\cline{3-6}&&7&13.7129943488902&&\\\hline\multirow{13}{*}{8}&\multirow{13}{*}{2}&1&1.271706087581433&1.670954477097469&\\\cline{3-6}&&2&2.383355863629706&1.662783111291984&\\\cline{3-6}&&3&2.092100093244703&-2.803377889217483&\\\cline{3-6}&&4&1.244313600108916&-3.065339745176996&\\\cline{3-6}&&5&2.334010482862648&4.308939906210139&\\\cline{3-6}&&6&-2.979846773221015&1.57827840091033&\\\cline{3-6}&&7&4.370551169701395&1.651842067444984&\\\cline{3-6}&&8&5.648765724234609&-2.237015855424188&\\\cline{3-6}&&9&1.269930725213365&2.390822553996926&\\\cline{3-6}&&10&-1.947287265389223&-4.491074032089162&\\\cline{3-6}&&11&-2.520489596445147&3.031731530022549&\\\cline{3-6}&&12&4.397977764490689&1.266495115558148&\\\cline{3-6}&&13&1.271475306333206&
 1.271475306333206&\\&&&+2.202304943896206$i$&-2.202304943896206$i$&\\\hline\multirow{27}{*}{8}&\multirow{27}{*}{3}&1&2.40873955875332&2.098628559113852&2.098628559113853\\&&&&-3.660393653953999$i$&+3.660393653953999$i$\\\cline{3-6}&&2&1.282713105358393&2.554950754236945&1.70924265930685\\\cline{3-6}&&3&2.283630889407069&2.28363088940707&1.664985843065564\\&&&-4.00022304697502$i$&+4.00022304697502$i$&\\\cline{3-6}&&4&5.529918109991038&1.51881768033327&1.518817680333269\\&&&&+2.631616151633382$i$&-2.631616151633382$i$\\\cline{3-6}&&5&7.414183066616086&-2.862682200630386&3.198078127975502\\\cline{3-6}&&6&-5.511247175838946&3.650488595458813&2.240655749924518\\\cline{3-6}&&7&3.340653884272446&-2.14255770445547&3.340653884272446\\&&&+5.919991271163282$i$&&-5.919991271163282$i$\\\cline{3-6}&&8&-5.514779752344294&-2.023919611315855&10.8777019433327\\\cline{3-6}&&9&1.68440705741468&5.423847662867905&2.467651140153592\\\cline{3-6}&&10&1.213531067160599&1.213531067160599&-10.489730959
 88394\\&&&-2.101911443901849$i$&+2.101911443901849$i$&\\\cline{3-6}&&11&-6.301965974660797&3.766071521088686&1.26000911004235\\\cline{3-6}&&12&4.885235830838998&-2.183128543959034&-9.430910891448477\\\cline{3-6}&&13&2.060403901016193&-2.658376699449247&-12.74552761325145\\\cline{3-6}&&14&1.278043319153198&1.693253474753915&5.596399575909029\\\cline{3-6}&&15&2.149598643270218&7.839163284765999&-3.330352881033191\\\cline{3-6}&&16&-11.81310452125254&-2.420711226941534&2.927550314738232\\\cline{3-6}&&17&-8.654318296711313&1.652191871107661&1.266379443459761\\\cline{3-6}&&18&-18.08314271073251&-1.91571801974974&-4.226460223694949\\\cline{3-6}&&19&1.269639528011694&2.349438781357117&2.349438781357117\\&&&&-4.1223176079197$i$&+4.1223176079197$i$\\\cline{3-6}&&20&5.471483838358137&2.476041590250572&1.276150039143275\\\cline{3-6}&&21&8.069263317405973&-3.594881640653509&1.599026959755484\\\cline{3-6}&&22&1.240656230548367&-2.889477062652692&-13.39895741005542\\\cline{3-6}&&23&2.30700
 4865853058&-7.830240072884703&1.642950953716392\\\cline{3-6}&&24&1.264378468655506&2.315531519097134&-8.061533718602336\\\cline{3-6}&&25&1.628626279736931&-6.078116022520112&3.732050807568877\\\cline{3-6}&&26&1.250953755245463&8.175846591759141&-3.714015014705677\\\cline{3-6}&&27&-2.812486479833951&-13.19173425438937&1.56692995698348\\\hline\end{tabular}
\caption{\small The admissible solutions of the XXZ Bethe equations
   (\ref{BAEx}) with $q=e^{i\pi/3}$ and $N=8$.} \label{table:XXZp3sltns2}
\end{table}

\begin{table}
\tiny \centering
\begin{tabular}{|c|c|c| c c c|}\hline
N&M&number  &$x_{1}$&$x_{2}$&$x_{3}$\\\hline
\multirow{1}{*}{2}&\multirow{1}{*}{1}&1&2.414213562373095&&\\\hline
\multirow{2}{*}{3}&\multirow{2}{*}{1}&1&1.628626279736931&&\\\cline{3-6}&&2&6.078116022520107&&\\\hline
\multirow{2}{*}{4}&\multirow{2}{*}{1}&1&2.414213562373095&&\\\cline{3-6}&&2&1.414213562373095&&\\\hline
\multirow{3}{*}{4}&\multirow{3}{*}{2}&1&1.45314130298278&3.20337817632093&\\\cline{3-6}&&2&2.668693617506732&2.668693617506733&\\&&&-3.09600332629458$i$&+3.09600332629458$i$&\\\hline
\multirow{4}{*}{5}&\multirow{4}{*}{1}&1&1.311033025558219&&\\\cline{3-6}&&2&1.861000175046023&&\\\cline{3-6}&&3&-8.277536750907217&&\\\cline{3-6}&&4&3.652418494292248&&\\\hline
\multirow{5}{*}{5}&\multirow{5}{*}{2}&1&1.339710835264441&2.046891922279901&\\\cline{3-6}&&2&7.305968916122279&1.320324940604242&\\\cline{3-6}&&3&1.91282417422196&6.940872062317657&\\\cline{3-6}&&4&2.064510526379247&2.064510526379247&\\&&&-2.02993224151061$i$&+2.02993224151061$i$&\\\hline
\multirow{5}{*}{6}&\multirow{5}{*}{1}&1&1.249688897773919&&\\\cline{3-6}&&2&1.628626279736931&&\\\cline{3-6}&&3&2.414213562373095&&\\\cline{3-6}&&4&-4.663902460147015&&\\\cline{3-6}&&5&6.078116022520107&&\\\hline
\multirow{5}{*}{6}&\multirow{5}{*}{2}&1&1.71311584242839&1.269294065270789&\\\cline{3-6}&&2&1.683266313484753&2.811927386738288&\\\cline{3-6}&&3&1.554024547007536&1.554024547007536&\\&&&+1.554715716529629$i$&-1.554715716529629$i$&\\\cline{3-6}&&4&2.846518894544238&1.263542099556734&\\\hline
\multirow{7}{*}{6}&\multirow{7}{*}{3}&1&1.262878926307629&3.075549320798162&3.075549320798161\\&&&&+3.542806348465251$i$&-3.542806348465251$i$\\\cline{3-6}&&2&1.771351991900205&3.826157503762062&1.281594555446714\\\cline{3-6}&&3&3.211911541809374&2.26937708343367&2.269377083433669\\&&&&+2.360695277629578$i$&-2.360695277629579$i$\\\cline{3-6}&&4&2.955672286788613&2.955672286788612&1.686278114645637\\&&&+3.365585528895605$i$&-3.365585528895605$i$&\\
\hline
\multirow{6}{*}{7}&\multirow{6}{*}{1}&1&1.208825798344529&&\\\cline{3-6}&&2&1.498359634541009&&\\\cline{3-6}&&3&1.986520161938681&&\\\cline{3-6}&&4&3.161528726585875&&\\\cline{3-6}&&5&-3.45462565368127&&\\\cline{3-6}&&6&13.29830956633742&&\\\hline\multirow{14}{*}{7}&\multirow{14}{*}{2}&1&1.222821216951139&1.547297967647689&\\\cline{3-6}&&2&1.220472188780732&2.150322976591122&\\\cline{3-6}&&3&2.140605616819436&1.536617282908898&\\\cline{3-6}&&4&2.067053540613296&4.01059689676756&\\\cline{3-6}&&5&2.879021158753284&2.879021158753283&\\&&&+2.956127592936119$i$&-2.956127592936119$i$&\\\cline{3-6}&&6&4.108663475752145&1.520996638277984&\\\cline{3-6}&&7&4.148660960002176&1.215637366932319&\\\cline{3-6}&&8&6.571169436423152&-4.94785739884573&\\\cline{3-6}&&9&-7.468286858137201&2.959302735558985&\\\cline{3-6}&&10&1.331990680492124&1.331990680492124&\\&&&+1.331978961361137$i$&-1.331978961361137$i$&\\\cline{3-6}&&11&-8.92688858805227&1.488188528857897&\\\cline{3-6}&&12&1.205729430210374
 &-9.095642759167061&\\\cline{3-6}&&13&1.951628400529481&-8.521599164762728&\\\cline{3-6}&&14&-16.46493396710602&-3.167610405963212&\\\hline
\multirow{8}{*}{7}&\multirow{8}{*}{3}&1&2.29407542825786&2.29407542825786&1.223807335674004\\&&&-2.262286996783694$i$&+2.262286996783694$i$&\\\cline{3-6}&&2&1.591961830774357&2.387325715651191&1.235084659761902\\\cline{3-6}&&3&2.208206873312259&2.20820687331226&1.552672568630353\\&&&-2.183599412028467$i$&+2.183599412028467$i$&\\\cline{3-6}&&4&2.258703430918202&1.923901682011331&1.923901682011331\\&&&&-1.916321113803041$i$&+1.91632111380304$i$\\\cline{3-6}&&5&1.226752347088832&8.364492571528938&1.561591283662463\\\cline{3-6}&&6&1.22436156832076&2.214765473037293&7.893846799575592\\\cline{3-6}&&7&1.399715240410388&9.104198650298802&1.399715240410387\\&&&+1.399671218147326$i$&&-1.399671218147326$i$\\\cline{3-6}&&8&2.204487438555672&7.745185417863576&1.550608398126798\\\hline
\end{tabular}
\caption{\small The admissible solutions of the XXZ Bethe equations
   (\ref{BAEx}) with $q=e^{i\pi/4}$ up to $N=7$.} \label{table:XXZp4sltns1}
\end{table}

\begin{table}
\tiny \centering
\begin{tabular}{|c|c|c| c c c|}\hline
N&M&number  &$x_{1}$&$x_{2}$&$x_{3}$\\\hline
\multirow{6}{*}{8}&\multirow{6}{*}{1}&1&1.179580427103275&&\\\cline{3-6}&&2&1.414213562373095&&\\\cline{3-6}&&3&1.765366864730179&&\\\cline{3-6}&&4&2.414213562373095&&\\\cline{3-6}&&5&-2.847759065022574&&\\\cline{3-6}&&6&4.261972627395668&&\\\hline\multirow{20}{*}{8}&\multirow{20}{*}{2}&1&1.446403950334763&1.190016384385456&\\\cline{3-6}&&2&1.852238654657947&1.188864345375984&\\\cline{3-6}&&3&2.514682127062517&6.295143655976973&\\\cline{3-6}&&4&2.307514146628479&-4.564170333529112&\\\cline{3-6}&&5&-4.028001605777415&3.623598033160472&\\\cline{3-6}&&6&5.426177490883765&5.42617749088376&\\&&&+6.696352141994065$i$&-6.696352141994067$i$&\\\cline{3-6}&&7&-8.475234192916405&-2.578922392281056&\\\cline{3-6}&&8&9.913152718172503&-3.206182813379592&\\\cline{3-6}&&9&2.677671679899443&1.82428114777186&\\\cline{3-6}&&10&1.18672458699157&2.711553564494324&\\\cline{3-6}&&11&1.43539240503802&2.700777985799591&\\\cline{3-6}&&12&2.237470143268887&2.237470143268887&\\&&&+2.232703480148387$i$&-
 2.232703480148388$i$&\\\cline{3-6}&&13&1.199443547942262&1.199443547942261&\\&&&+1.199443691222013$i$&-1.199443691222013$i$&\\\cline{3-6}&&14&1.441546045187176&1.848322534982018&\\\cline{3-6}&&15&1.423573997535758&6.679481637950619&\\\cline{3-6}&&16&-4.822615261912324&1.733667260375862&\\\cline{3-6}&&17&1.402547159199412&-4.948633639651279&\\\cline{3-6}&&18&6.730585662140516&1.182717881167556&\\\cline{3-6}&&19&-5.007962765238115&1.175606675111065&\\\cline{3-6}&&20&6.565048430866063&1.791824885481342&\\\hline\multirow{8}{*}{8}&\multirow{8}{*}{3}&1&1.933808178839618&1.545539874804185&1.545539874804185\\&&&&-1.545666045059988$i$&+1.545666045059988$i$\\\cline{3-6}&&2&1.347234359171865&1.347234359171865&3.304419901978651\\&&&-1.347240969937556$i$&+1.347240969937556$i$&\\\cline{3-6}&&3&1.468276599483929&3.220895697334547&1.19694970312494\\\cline{3-6}&&4&1.651830400488286&1.651830400488287&1.464153355718869\\&&&-1.6521877840048$i$&+1.6521877840048$i$&\\\cline{3-6}&&5&1.200048659530
 919&1.961718914285269&1.477882030590439\\\cline{3-6}&&6&1.462678390951383&3.155365989103904&1.916232292792105\\\cline{3-6}&&7&1.195184858596888&1.694235432921724&1.694235432921724\\&&&&-1.694739206592311$i$&+1.694739206592311$i$\\\cline{3-6}&&8&3.174113422492026&1.92094315096371&1.195647766853818\\\hline
\end{tabular}
\caption{\small The admissible solutions of the XXZ Bethe equations
   (\ref{BAEx}) with $q=e^{i\pi/4}$ and $N=8$.} \label{table:XXZp4sltns2}
\end{table}

\begin{table}
\tiny \centering
\begin{tabular}{|c|c|c| c c c|}\hline
N&M&number  &$x_{1}$&$x_{2}$&$x_{3}$\\\hline
\multirow{1}{*}{2}&\multirow{1}{*}{1}&1&1.962610505505151&&\\\hline
\multirow{2}{*}{3}&\multirow{2}{*}{1}&1&1.461818651603003&&\\\cline{3-6}&&2&3.574329190217505&&\\\hline
\multirow{3}{*}{4}&\multirow{3}{*}{1}&1&1.96261050550515&&\\\cline{3-6}&&2&1.311033025558219&&\\\cline{3-6}&&3&8.27753675090721&&\\\hline
\multirow{3}{*}{4}&\multirow{3}{*}{2}&1&1.344028591243017&2.447657661838913&\\\cline{3-6}&&2&2.189055410994626&2.189055410994627&\\&&&-1.71493838744667$i$&+1.71493838744667$i$&\\\hline
\multirow{3}{*}{5}&\multirow{3}{*}{1}&1&1.23606797749979&&\\\cline{3-6}&&2&1.618033988749895&&\\\cline{3-6}&&3&2.618033988749894&&\\\hline
\multirow{7}{*}{5}&\multirow{7}{*}{2}&1&1.749592331415895&1.259572013945332&\\\cline{3-6}&&2&1.667690311962022&3.974545391101193&\\\cline{3-6}&&3&1.724293088189099&1.724293088189099&\\&&&+1.246431541912111$i$&-1.246431541912111$i$&\\\cline{3-6}&&4&4.116005110398541&1.246284665265577&\\\cline{3-6}&&5&4.142310683532282&4.142310683532279&\\&&&+4.35688110427138$i$&-4.356881104271381$i$&\\\hline
\multirow{5}{*}{6}&\multirow{5}{*}{1}&1&1.190729200383194&&\\\cline{3-6}&&2&1.461818651603003&&\\\cline{3-6}&&3&1.96261050550515&&\\\cline{3-6}&&4&3.574329190217505&&\\\cline{3-6}&&5&-10.40659374764855&&\\\hline
\multirow{10}{*}{6}&\multirow{10}{*}{2}&1&1.206790520429557&1.524710031541862&\\\cline{3-6}&&2&2.231459158071619&1.202712882967582&\\\cline{3-6}&&3&2.007446763339642&8.411463578743241&\\\cline{3-6}&&4&1.505088894154398&2.211779917490494&\\\cline{3-6}&&5&3.391097834451792&3.391097834451791&\\&&&+2.137260009634191$i$&-2.137260009634192$i$&\\\cline{3-6}&&6&1.429452052607576&1.429452052607576&\\&&&+1.038668851613124$i$&-1.038668851613125$i$&\\\cline{3-6}&&7&1.193470251543278&9.123706421008574&\\\cline{3-6}&&8&8.943547047241228&1.471698013592016&\\\hline
\multirow{9}{*}{6}&\multirow{9}{*}{3}&1&1.874104285363227&1.874104285363228&2.589407182559046\\&&&-1.378223914748844$i$&+1.378223914748843$i$&\\\cline{3-6}&&2&1.205338345735675&2.435006504125385&2.435006504125385\\&&&&+1.895623247953007$i$&-1.895623247953007$i$\\\cline{3-6}&&3&4.792019859888095&1.147122809503764&1.147122809503762\\&&&&+4.191977909831017$i$&-4.191977909831016$i$\\\cline{3-6}&&4&1.217923140935344&2.813376407957134&1.572774650848652\\\cline{3-6}&&5&2.352950008994888&2.352950008994887&1.52038626323149\\&&&+1.817348862908779$i$&-1.81734886290878$i$&\\\hline
\multirow{6}{*}{7}&\multirow{6}{*}{1}&1&1.160202334045279&&\\\cline{3-6}&&2&1.37099723775096&&\\\cline{3-6}&&3&1.699473589729839&&\\\cline{3-6}&&4&2.375165198312843&&\\\cline{3-6}&&5&-5.721812045187025&&\\\cline{3-6}&&6&5.148221975956501&&\\\hline\multirow{8}{*}{7}&\multirow{8}{*}{2}&1&1.408432222064447&1.171698991051096&\\\cline{3-6}&&2&1.401167728417128&1.805963656505584&\\\cline{3-6}&&3&1.812019057106945&1.170007362023291&\\\cline{3-6}&&4&1.279250449479749&1.279250449479748&\\&&&+0.9294284209179594$i$&-0.9294284209179595$i$&\\\cline{3-6}&&5&2.170715518398742&2.170715518398741&\\&&&+1.586182712241336$i$&-1.586182712241336$i$&\\\cline{3-6}&&6&1.762389509249143&2.822111962624334&\\\cline{3-6}&&7&1.166632159289219&2.888620018910967&\\\cline{3-6}&&8&2.868965063692831&1.390952988340999&\\\hline\multirow{13}{*}{7}&\multirow{13}{*}{3}&1&2.618379654246318&3.748323531486734&3.748323531486732\\&&&&+3.748017767598491$i$&-3.748017767598492$i$\\\cline{3-6}&&2&0.9827111074314028&3.402279
 883618681&0.982711107431404\\&&&-3.262273894557891$i$&&+3.262273894557891$i$\\\cline{3-6}&&3&1.44412440163796&1.975097530601085&1.182185195050447\\\cline{3-6}&&4&1.639956241140929&1.639956241140929&1.92399300966388\\&&&+1.19052263024356$i$&-1.19052263024356$i$&\\\cline{3-6}&&5&1.422694691325217&1.811547625097516&1.811547625097515\\&&&&+1.312430925490274$i$&-1.312430925490274$i$\\\cline{3-6}&&6&1.865424819629057&1.865424819629057&1.175408810971523\\&&&+1.350217351171199$i$&-1.3502173511712$i$&\\\cline{3-6}&&7&4.278808762679677&4.278808762679675&1.735950022329603\\&&&+4.435213551496066$i$&-4.435213551496067$i$&\\\cline{3-6}&&8&4.482031924050006&4.482031924050003&1.163787451697995\\&&&+4.683990539341568$i$&-4.683990539341569$i$&\\\cline{3-6}&&9&1.176249499108818&4.567826614400596&1.423871663317115\\\cline{3-6}&&10&1.174464705246947&1.872470411706964&4.389889034358894\\\cline{3-6}&&11&1.351963489253257&4.883627945846047&1.351963489253257\\&&&+0.9822482229626871$i$&&-0.9822482229
 626871$i$\\\cline{3-6}&&12&1.382220257632069&4.424719229653134&4.424719229653131\\&&&&+4.614269329027579$i$&-4.61426932902758$i$\\\cline{3-6}&&13&1.865730317958417&4.331809157967312&1.416090213898351\\\hline
\end{tabular}
\caption{\small The admissible solutions of the XXZ Bethe equations
   (\ref{BAEx}) with $q=e^{i\pi/5}$ up to $N=7$.} \label{table:XXZp5sltns1}
\end{table}

\begin{table}
\tiny \centering
\begin{tabular}{|c|c|c| c c c|}\hline
N&M&number  &$x_{1}$&$x_{2}$&$x_{3}$\\\hline
\multirow{7}{*}{8}&\multirow{7}{*}{1}&1&1.138192552062957&&\\\cline{3-6}&&2&1.311033025558219&&\\\cline{3-6}&&3&1.96261050550515&&\\\cline{3-6}&&4&1.554619032420834&&\\\cline{3-6}&&5&2.893146291465051&&\\\cline{3-6}&&6&-4.157155527220437&&\\\cline{3-6}&&7&8.27753675090721&&\\\hline\multirow{20}{*}{8}&\multirow{20}{*}{2}&1&1.336088158806443&1.146789909509786&\\\cline{3-6}&&2&1.614712957660528&1.332716767162854&\\\cline{3-6}&&3&2.129163003251931&1.599504552643429&\\\cline{3-6}&&4&1.328550709738535&2.142233338886835&\\\cline{3-6}&&5&1.145952463660923&1.617263878227611&\\\cline{3-6}&&6&2.148429235497258&1.144427236851177&\\\cline{3-6}&&7&5.970673342816859&-6.807556417171186&\\\cline{3-6}&&8&1.186193082020218&1.186193082020218&\\&&&+0.8618197338868376$i$&-0.8618197338868376$i$&\\\cline{3-6}&&9&3.700975871220594&2.044413881921601&\\\cline{3-6}&&10&3.22254557647494&3.22254557647494&\\&&&+2.442247793000595$i$&-2.442247793000596$i$&\\\cline{3-6}&&11&1.816361993233109&1.816361993233108
 &\\&&&+1.319263515796331$i$&-1.319263515796331$i$&\\\cline{3-6}&&12&1.580119483569546&3.800019441082347&\\\cline{3-6}&&13&1.308213900461231&-10.8884292670433&\\\cline{3-6}&&14&-10.96882884583839&1.137180644939945&\\\cline{3-6}&&15&1.941451195793101&-10.35085068475084&\\\cline{3-6}&&16&3.844192024006444&1.320876365901175&\\\cline{3-6}&&17&-9.448193511791008&2.79024143505726&\\\cline{3-6}&&18&-10.71635263086061&1.547521808575225&\\\cline{3-6}&&19&1.141680918344861&3.864338139442635&\\\cline{3-6}&&20&-20.75089982455659&-3.747556189954495&\\\hline
\multirow{21}{*}{8}&\multirow{21}{*}{3}&1&1.152911608122782&1.521182841520322&1.521182841520321\\&&&&+1.105269705773082$i$&-1.105269705773082$i$\\\cline{3-6}&&2&1.689476731560921&1.420511692843986&1.420511692843985\\&&&&+1.032076474649996$i$&-1.032076474649996$i$\\\cline{3-6}&&3&2.116551646515748&3.190321178102575&3.190321178102574\\&&&&+2.114684828886933$i$&-2.114684828886935$i$\\\cline{3-6}&&4&1.152031745303854&1.670604849464544&2.431584410448484\\\cline{3-6}&&5&1.49131548027285&1.49131548027285&1.355505247355462\\&&&+1.083548228240326$i$&-1.083548228240326$i$&\\\cline{3-6}&&6&1.153006146461386&2.457631476354184&1.354649410334471\\\cline{3-6}&&7&1.155264412504946&1.696786869605106&1.361288361070377\\\cline{3-6}&&8&1.301230331780993&1.301230331780993&2.520770703624886\\&&&+0.945400143152111$i$&-0.945400143152111$i$&\\\cline{3-6}&&9&2.767507419138274&0.8536816467160973&0.8536816467160964\\&&&&+2.639644349871746$i$&-2.639644349871746$i$\\\cline{3-6}&&10&1.667446266251631&2.420
 923118070947&1.350646860029378\\\cline{3-6}&&11&3.60404301030639&3.604043010306389&1.14351105004767\\&&&+2.298115335328595$i$&-2.298115335328597$i$&\\\cline{3-6}&&12&2.17970199500867&8.743528394000164&1.611362131547616\\\cline{3-6}&&13&9.641933706319058&1.629607621987226&1.147392392511266\\\cline{3-6}&&14&3.466847839330259&3.466847839330258&1.595747808557541\\&&&+2.237018668690406$i$&-2.237018668690407$i$&\\\cline{3-6}&&15&1.203999000879017&10.7917885753781&1.203999000879017\\&&&+0.8747565077890832$i$&&-0.8747565077890832$i$\\\cline{3-6}&&16&1.91597854331014&9.300125780741034&1.91597854331014\\&&&+1.390808610142858$i$&&-1.390808610142858$i$\\\cline{3-6}&&17&1.340380214566179&9.874749723055375&1.148237393071579\\\cline{3-6}&&18&2.199876642321688&9.032424520308002&1.145845769779947\\\cline{3-6}&&19&1.332719524956927&2.193418356684709&8.945501899172173\\\cline{3-6}&&20&1.336967053063209&9.546671796790799&1.627009052110716\\\cline{3-6}&&21&1.326274265652015&3.563336816528149&3.5
 63336816528148\\&&&&+2.279904684704862$i$&-2.279904684704864$i$\\\hline
\end{tabular}
\caption{\small The admissible solutions of the XXZ Bethe equations
   (\ref{BAEx}) with $q=e^{i\pi/5}$ and $N=8$.} \label{table:XXZp5sltns2}
\end{table}

\clearpage


\providecommand{\href}[2]{#2}\begingroup\raggedright\endgroup

\end{document}